\documentclass[10pt,aps,prb,twocolumn,amsmath,amssymb,superscriptaddress]{revtex4-1}
\usepackage{amsmath}
\usepackage{graphicx}
\usepackage{color}
\usepackage{multirow}
\usepackage[caption = false]{subfig}
\usepackage{array}
\usepackage{threeparttable}
\usepackage{mhchem}
\usepackage{mathtools}
\usepackage{siunitx}
\usepackage{centernot}
\usepackage{tikz}
\tikzset{font={\fontsize{11pt}{12}\selectfont}}
\usetikzlibrary{shapes,arrows}
\usepackage{lipsum}
\makeatletter

\renewcommand*{\p@subsection}{}

\renewcommand*{\p@subsubsection}{}
\makeatother

\usepackage{hyperref}
\hypersetup{
	linkcolor=blue,          % color of internal links (change box color with linkbordercolor)
	citecolor=blue,        % color of links to bibliography
	urlcolor=blue,           % color of external links
}

\newcommand{\ket}[1]{|#1\rangle}

\newcommand{\inner}[2]{\langle#1|#2\rangle}

\newcommand{\expecth}[3]{\langle#1|#2|#3\rangle}

\begin{document}
\title{Selected configuration interaction wave functions in phaseless auxiliary field quantum Monte Carlo}

\author{Ankit Mahajan}
\email{ankitmahajan76@gmail.com}
\affiliation{Department of Chemistry, University of Colorado, Boulder, CO 80302, USA}

\author{Joonho Lee}
\affiliation{Department of Chemistry, Columbia University, New York, New York 10027, USA}
\author{Sandeep Sharma}
\email{sanshar@gmail.com}
\affiliation{Department of Chemistry, University of Colorado, Boulder, CO 80302, USA}
\begin{abstract}
We present efficient algorithms for using selected configuration interaction (sCI) trial wave functions in phaseless auxiliary field quantum Monte Carlo (ph-AFQMC). These advancements, geared towards optimizing computational performance for longer CI expansions, allow us to use up to a million configurations feasibly in the trial state for ph-AFQMC. In one example, we found the cost of ph-AFQMC per sample to increase only by a factor of about 3 for a calculation with \(10^4\) configurations compared to that with a single one, demonstrating the tiny computational overhead due to a longer expansion. This favorable scaling allows us to study the systematic convergence of the phaseless bias in AFQMC calculations with an increasing number of configurations and provides a means to gauge the accuracy of ph-AFQMC with other trial states. We also show how the scalability issues of sCI trial states for large system sizes could be mitigated by restricting them to a moderately sized orbital active space and leveraging the near-cancellation of out of active space phaseless errors.
\end{abstract}
\maketitle

\section{Introduction}

Quantum Monte Carlo (QMC) is a powerful tool in our arsenal to tackle the quantum many-body problem.\cite{kalos1974helium,ceperley1977monte,ceperley1980ground,nightingale1998quantum,foulkes2001quantum,booth2013towards,becca2017quantum} Among various QMC approaches, phaseless auxiliary-field quantum Monte Carlo (ph-AFQMC)\cite{zhang2003quantum} has emerged as an accurate and efficient method. While originally from the condensed matter community (therein usually referred to as constrained-path AFQMC),\cite{fahy1990positive,zhang1995constrained,zhang1997constrained} ph-AFQMC has gained popularity in chemistry in recent years.\cite{al2006auxiliary,al2007bond,suewattana2007phaseless,purwanto2015auxiliary,motta2018ab,hao2018accurate,motta2019efficient,shee2019singlet,shee2019achieving,williams2020direct,lee2020stochastic,lee2020utilizing,shi2021some} The accuracy and scalability of ph-AFQMC are determined in large part by the choice of the trial wave function. The use of a trial wave function becomes necessary for retaining statistical efficiency (in sample complexity) to control the fermionic phase (or sign) problem\cite{loh1990sign,troyer2005computational}. The constraint imposed to control the phase problem is called the phaseless approximation\cite{zhang2003quantum}. When the trial wave function approaches the exact ground state, the corresponding ph-AFQMC energy tends to the  ground state energy. The commonly used trial wave function for ph-AFQMC is the broken-symmetry Hartree-Fock (HF) wave function. It scales as $O(N^3)$ to obtain the trial wave function and $O(N^5)$ to perform the ph-AFQMC calculation with the trial to obtain energy for a fixed statistical error, where \(N\) is the system size. Beyond HF, one may try to use a single determinant trial using approximate Bruckner orbitals\cite{lee2020utilizing}, which essentially keeps the same computational scaling for ph-AFQMC. While single determinant trial wave functions are attractive due to their scalability, their accuracy can be limited and questionable in many examples\cite{al2007bond,williams2020direct,mahajan2021taming}.

On the other hand, there has been a flurry of developments in selected configuration interaction (sCI) methods in the last few years.\cite{giner2013using,evangelista2014adaptive,Holmes2016b,tubman2016deterministic} Despite their steep (formally exponential) scaling with system size, sCI methods have increasingly been employed to study moderately sized systems and perform large active space correlated calculations. These methods have the capability of generating systematically more accurate approximations to the state of interest by increasing the number of configurations judiciously. This attractive feature makes sCI wave functions natural candidates for trial states in projection QMC methods. Based on significant advances in real-space QMC algorithms with such trial states,\cite{filippi2016simple,Assaraf2017} several papers have already reported calculations with long sCI expansions in diffusion Monte Carlo.\cite{dash2018perturbatively,pineda2019excited,benali2020toward} There has also been some recent progress for using them in ph-AFQMC,\cite{shee2018phaseless,williams2020direct,shi2021some} but the commonly used algorithm based on the Sheman-Morrison-Woodbury identity has an inherent computational bottleneck with the local energy evaluation scaling as $O(N_c N^4)$, where $N_c$ is the number of determinants. This steeper scaling compared to real-space methods is due to the Gaussian orbital representation of the electron repulsion interaction. Furthermore, the force bias evaluation scales as $O(N_c N^3)$, which can be expensive since it is performed once every time step. Due to these, it has been challenging to scale up the ph-AFQMC calculations beyond 100-1000 configurations.

In this work, we propose to use one of the sCI methods, heat-bath CI (HCI)\cite{Holmes2016b,ShaHolUmr,smith2017cheap}, to generate a large trial with up to $10^{6}$ configurations. We then combine several algorithmic advances made by two of us\cite{mahajan2020efficient,mahajan2021taming} using the generalized Wick's theorem and make further improvements to accelerate ph-AFQMC calculations with a large trial. In particular, in our new algorithm, the local energy evaluation scales as $O(N_c N + N^4)$, and the force bias evaluation scales as $O(N_c + N^3)$. To alleviate the scaling issues of sCI wave functions, we study the efficacy of active space trial states in calculating energy differences and properties. The active space size is another variable to converge the phaseless error systematically in larger systems while keeping the calculation cost manageable. We use the new algorithm for ph-AFQMC with HCI trials to investigate the behavior of the phaseless bias as a function of the number of configurations in the trial for several challenging systems.

This paper is organized as follows: we first review the ph-AFQMC algorithm (Section \ref{sec:phaseless_review}), we then discuss how once can drastically speed up the evaluation of force bias and local energy of sCI trials using the generalized Wick's theorem (Section \ref{sec:sci_trial}), we present results that show how phaseless errors in the ground state energy and dipole moments change as a function of the number of determinants in hydrogen chains, transition metal oxides, and a few small molecules (Section \ref{sec:results}), and we finally conclude (Section \ref{sec:conclusion}).

\section{Phaseless auxiliary field quantum Monte Carlo}\label{sec:phaseless_review}
We briefly summarize the procedure for phaseless AFQMC here and refer the reader to reference \onlinecite{motta2018ab} for more details. Consider the \textit{ab initio} Hamiltonian given by
\begin{equation}
	\hat{H} = \sum_{ij} h_{ij}\hat{a}_i^{\dagger}\hat{a}_j + \frac{1}{2}\sum_{\gamma}\left(\sum_{ij}L^{\gamma}_{ij}\hat{a}_i^{\dagger}\hat{a}_j\right)^2,\label{eq:ham}
\end{equation}
where \(h_{ij}\) are one-electron integrals and \(L^{\gamma}_{ij}\) are Cholesky decomposed two-electron integrals in an orthonormal orbital basis. We will use letters \(N\), \(M\), and \(X\) to denote the number of electrons, number of orbitals, and number of Cholesky vectors, respectively. We note that in chemical systems, empirically, \(X \sim O(M)\) with a proportionality constant usually smaller than 10. AFQMC uses the exponential form of the projector to converge to the ground state of the Hamiltonian as
\begin{equation}
   e^{-\tau \hat{H}}\ket{\psi_I} \xrightarrow{\tau\rightarrow\infty}\ket{\Psi_0},
\end{equation}
where \(\tau\) is the imaginary time, \(\ket{\Psi_0}\) is the ground state, and \(\ket{\psi_I}\) is an initial state such that \(\inner{\psi_I}{\psi_0}\neq 0\). Using the Hubbard-Stratonovic transform, the short-time exponential projector can be written as
\begin{equation}
   e^{-\Delta\tau\hat{H}} = \int d \mathbf{x} p(\mathbf{x})\hat{B}(\mathbf{x}),\label{eq:prop}
\end{equation}
where \(\mathbf{x}\) is the vector of auxiliary fields (one scalar field per Cholesky component), \(p(\mathbf{x})\) is the standard normal Gaussian distribution of the auxiliary fields, and \(\hat{B}(\mathbf{x})\) is a complex propagator given by the exponential of a one-body operator. Due to Thouless' theorem, \(\hat{B}(\mathbf{x})\) acts on a Slater determinant \(\ket{\phi}\) as
\begin{equation}
   \hat{B}(\mathbf{x})\ket{\phi} = \ket{\phi(\mathbf{x})},
\end{equation}
where \(\ket{\phi(\mathbf{x})}\) is another Slater determinant obtained by a linear transformation of the orbitals in \(\ket{\phi}\). By importance sampling the auxiliary fields in equation \ref{eq:prop} and applying the short time propagator sufficiently many times, we get a stochastic representation of the ground state wave function in the long time limit as
\begin{equation}
   \ket{\Psi_0} \propto \sum_i w_i \frac{\ket{\phi_i}}{\inner{\psi_T}{\phi_i}},
\end{equation}
where \(w_i\) are weights, \(\ket{\phi_i}\) are Slater determinants with complex orbitals obtained from the orbital transformations sampled during propagation, and \(\ket{\psi_T}\) is the trial state used for importance sampling. More accurate trial states lead to less noisy weights. We use the hybrid approximation in this paper, which avoids the expensive calculation of local energy at each propagation step. Importance sampling in AFQMC involves shifting the sampled auxiliary fields by the force bias given as
\begin{equation}
   \bar{x}_{\gamma} = -\sqrt{\Delta\tau} \frac{\expecth{\psi_T}{\sum_{ij}L^{\gamma}_{ij}\hat{a}_i^{\dagger}\hat{a}_j}{\phi}}{\inner{\psi_T}{\phi}},\label{eq:fb}
\end{equation}
where \(\ket{\phi}\) is the walker. The force bias can be thought of as a dynamic correction to the mean-field contour shift\cite{rom1997shifted} that vanishes as \(\Delta\tau \rightarrow 0\). Force bias is not enough, by itself, to control the large fluctuations stemming from the phase problem. The phaseless constraint can be used to overcome the phase problem at the expense of a systematic bias in the sampled wave function. The size of this phaseless bias is dictated by the accuracy of the trial state.

After an equilibration time, the energy of the sampled wave function can be measured as
\begin{equation}
   E \approx \frac{\sum_i w_i E_L(\phi_i)}{\sum_i w_i},
\end{equation}
where
\begin{equation}
   E_L(\phi_i) = \frac{\expecth{\psi_T}{\hat{H}}{\phi_i}}{\inner{\psi_T}{\phi_i}}\label{eq:eloc}
\end{equation}
is the local energy of the walker \(\ket{\phi_i}\). We note that this estimator has a zero-variance property i.e. in the limit that the trial state is the exact, ground state the variance of the estimator vanishes. In practice, this means that more accurate trial states lead to less noisy local energy values.

It is evident that more accurate trial states lead to less noisy simulations and smaller phaseless biases. But, in general, the use of more sophisticated trial states comes at the expense of greater computational cost. In the following section, we present efficient algorithms to calculate force bias and local energy of sCI trial states.

\section{Selected configuration interaction trial states} \label{sec:sci_trial}
A selected CI wave function is given by
\begin{equation}
   \ket{\psi_T} = \sum_{n}^{N_c}c_n \ket{\psi_n} = \sum_{n}^{N_c}c_n\prod_{\mu}^{k_n}\hat{a}^{\dagger}_{t_{n_{\mu}}}\hat{a}_{p_{n_{\mu}}}\ket{\psi_0},\label{eq:sci}
\end{equation}
where \(\ket{\psi_n}\) are configurations obtained by particle-hole excitations from the reference configuration \(\ket{\psi_0}\), \(N_c\) is the number of configurations in the expansion, \(c_n\) are real expansion coefficients, and \(k_n\) are the excitation ranks. We will use the indices \(p_{\mu}\) and \(t_{\mu}\) to denote occupied and virtual orbitals, respectively, whereas indices \(i, j, \dots\) will be used for general orbitals. The orthonormal orbital basis set used in the expansion can be chosen to be natural orbitals obtained from an HCI calculation or can be optimized using a self-consistent procedure. In the following, we use the same basis set to express the Hamiltonian in equation \ref{eq:ham} and, in general, all second-quantized operators refer to this basis. In our prior work, we have proposed efficient algorithms for using selected CI states in variational Monte Carlo\cite{mahajan2020efficient} and free projection AFQMC.\cite{mahajan2021taming} These focused on the calculation of local energy using the generalized Wick's theorem. Here, we discuss an algorithm for calculating the force bias required in ph-AFQMC and briefly summarize the calculation of local energy.

\subsection{Force bias}\label{sec:fb}
Calculation of the force bias for importance sampling is one of the computationally intensive parts of propagation in ph-AFMQC. Substituting the sCI trial state into the force bias expression in equation \ref{eq:fb}, we get
\begin{equation}
   \bar{x}_{\gamma} = -\sqrt{\Delta\tau}\frac{\sum_n\sum_{ij}c_nL^{\gamma}_{ij}\expecth{\psi_n}{\hat{a}_i^{\dagger}\hat{a}_j}{\phi}}{\sum_nc_n\inner{\psi_n}{\phi}}.\label{eq:msfb}
\end{equation}
The conventional way of evaluating the force bias proceeds by separately calculating the contribution of each configuration in the CI expansion using the Sherman-Morrison-Woodbury identity, resulting in a cost scaling of \(O(N_cNM + XNM)\).\cite{shee2018phaseless} Here we present an algorithm with cost scaling as \(O(N_c + XM^2)\) by essentially separating the sum over CI excitations from that over Cholesky integral indices in equation \ref{eq:msfb}.

We define the Green's function matrix for the reference configuration \(\ket{\psi_0}\) and the walker configuration \(\ket{\phi}\) as\cite{lowdin1955quantum,balian1969nonunitary}
\begin{equation}
   G^i_j = \frac{\expecth{\psi_0}{\hat{a}_i^{\dagger}\hat{a}_j}{\phi}}{\inner{\psi_0}{\phi}} = \left[\phi(\psi_0^{\dagger}\phi)^{-1}\psi_0^{\dagger}\right]^j_i,
\end{equation}
where \(\psi_0\) and \(\phi\) are the orbital coefficient matrices of the corresponding configurations, and superscripts and subscripts denote row and column indices, respectively. Note that since we work in the orbital basis of the CI expansion, \(\psi_0\) has a particularly simple form with all its columns being unit vectors. Thus the cost of calculating the Green's function in this basis scales as \(O(N^2M)\). For convenience, we also define the related quantity
\begin{equation}
   \mathcal{G}^i_j = -\frac{\expecth{\psi_0}{\hat{a}_j\hat{a}_i^{\dagger}}{\phi}}{\inner{\psi_0}{\phi}} = G^i_j - \delta^i_j.
\end{equation}
According to the generalized Wick's theorem\cite{balian1969nonunitary}, we have
\begin{equation}
   \frac{\expecth{\psi_0}{\prod_{\mu}^k \hat{a}^{\dagger}_{p_{\mu}}\hat{a}_{t_{\mu}}}{\phi}}{\inner{\psi_0}{\phi}} = \det\left(G^{\left\{p_{\mu}\right\}}_{\left\{t_{\mu}\right\}}\right),
\end{equation}
where the sets of indices \(\left\{p_{\mu}\right\}\) and \(\left\{t_{\mu}\right\}\) denote the \(k\times k\) slice of the \(G\) matrix, \(k\) being the rank of the excitation. This expression is obtained by taking pairwise contractions of the operators in the string of excitations according to the generalized Wick's theorem, with the determinant structure arising due to fermionic permutation parity factors.  Therefore, the denominator in equation \ref{eq:msfb} can be expressed as
\begin{equation}
   \frac{\inner{\psi_T}{\phi}}{\inner{\psi_0}{\phi}} = \sum_n c_n \det\left(G^{\left\{p_{n_{\mu}}\right\}}_{\left\{t_{n_{\mu}}\right\}}\right).
\end{equation}
The computational cost scaling of this overlap ratio is thus \(O(N_c)\) once the Green's function has been calculated. Since the Wick's theorem applies naturally to overlap ratios, we will find it convenient to express matrix elements as ratios with the reference overlap \(\inner{\psi_0}{\phi}\).

To evaluate the numerator of equation \ref{eq:msfb}, consider the matrix element ratio given by
\begin{equation}
   \frac{\expecth{\psi_n}{\hat{a}_i^{\dagger}\hat{a}_j}{\phi}}{\inner{\psi_0}{\phi}} = \frac{\expecth{\psi_0}{\left(\prod_{\mu}^k \hat{a}^{\dagger}_{p_{\mu}}\hat{a}_{t_{\mu}}\right)\hat{a}_i^{\dagger}\hat{a}_j}{\phi}}{\inner{\psi_0}{\phi}},
\end{equation}
where we have dropped the configuration index subscript (\(n\)) on the excitations for clarity. This ratio can also be evaluated using the generalized Wick's theorem and by choosing the appropriate order of operations one can achieve a significant reduction in the cost of calculating the numerator. To this end, we note that the pairwise contractions in the string of excitation operators can be divided into two groups: \(i\)) those without any contractions between CI excitations and \(\hat{a}_i^{\dagger}\hat{a}_j\), \(ii\)) those containing contractions between two of the CI operators and \(\hat{a}_i^{\dagger}\hat{a}_j\). Algebraically, the two terms are given as
 \begin{equation}
   \begin{split}
    &\frac{\expecth{\psi_0}{\left(\prod_{\mu}^{k} a^{\dagger}_{p_{\mu}}a_{t_{\mu}}\right)\hat{a}_i^{\dagger}\hat{a}_j}{\phi}}{\inner{\psi_0}{\phi}} = \det \begin{pmatrix}
			G^i_j & \mathcal{G}^i_{\left\{t_{\mu}\right\}}\\[1em]
			G^{\left\{p_{\mu}\right\}}_j & G^{\left\{p_{\mu}\right\}}_{\left\{t_{\mu}\right\}}\\
		\end{pmatrix}\\
		&\ = G^i_j \det\left(G^{\left\{p_{\mu}\right\}}_{\left\{t_{\mu}\right\}}\right)\\
		&\qquad\qquad\ + \sum_{\nu,\lambda}^k(-1)^{1+\nu+\lambda}G^{p_{\nu}}_j\mathcal{G}^i_{t_{\lambda}}\det\left(G^{\left\{p_{\mu}\right\}\symbol{92}p_{\nu}}_{\left\{t_{\mu}\right\}\symbol{92}t_{\lambda}}\right),
   \end{split}
\end{equation}
where on the first line the matrix of size \((k+1)\times (k+1)\) is written in a block form, and the notation \(\left\{p_{\mu}\right\}\symbol{92}p_{\nu}\) indicates the set of indices \(\left\{p_{\mu}\right\}\) excluding \(p_{\nu}\). Using this grouping of terms, we get for the numerator in equation \ref{eq:msfb}
\begin{equation}
  \begin{split}
    \sum_n&\sum_{ij}c_nL_{ij}^{\gamma}\frac{\expecth{\psi_n}{\hat{a}_i^{\dagger}\hat{a}_j}{\phi}}{\inner{\psi_0}{\phi}}\\
		&= \left(\sum_n c_n \det\left(G^{\left\{p_{\mu_n}\right\}}_{\left\{t_{\mu_n}\right\}}\right)\right)\left(\sum_{ij}L^{\gamma}_{ij}G^i_j\right)\\
		&\qquad\qquad\qquad\qquad+ \sum_{pt}I^t_p\left(\sum_{ij} L^{\gamma}_{ij}G^p_j\mathcal{G}^i_t\right),
  \end{split}\label{eq:fb_eff}
\end{equation}
where
\begin{equation}
   I^t_p = \sum_n c_n \sum_{\nu,\lambda}^{k_n}\delta_{p_{n_{\nu}},p}\delta_{t_{n_{\lambda}},t}(-1)^{1+\nu+\lambda}\det\left(G^{\left\{p_{n_{\mu}}\right\}\symbol{92}p_{n_{\nu}}}_{\left\{t_{n_{\mu}}\right\}\symbol{92}t_{n_{\lambda}}}\right)
\end{equation}
is an intermediate formed by summing over the CI configurations. The two terms in equation \ref{eq:fb_eff} are shown as tensor contraction diagrams in table \ref{tab:scaling} for the case of doubly excited CI configurations. Note that a selected CI state does not necessarily include all excitations of a given rank, but since it is not well suited for a tensor representation we use all doubly excited configurations in the table. In practice, the sum over CI excitations in table \ref{tab:scaling} is only performed over the configurations in the selected CI state. The first term has a cost scaling of \(O(N_c + XNM)\), while the second one using the intermediate has a scaling of \(O(N_c + XM^2)\). If the selected CI state is restricted to an active space of size \(A\), the overall cost can be reduced to \(O(N_c + XNM + XAM)\).

\begin{table*}
\caption{Summary of computational cost scaling of force bias and two-body local energy calculations (given the reference Green's function matrix \(G\)). SMW refers to overall cost scaling of the algorithm that makes use of the Sherman-Morrison-Woodbury formula as described in reference \onlinecite{shee2018phaseless}. The last columns shows tensor contraction diagrams for all distinct types of pairwise Wick's contractions using \textit{all} doubly excited configurations \(\sum c_{ptqu}\hat{a}_t^{\dagger}\hat{a}_p\hat{a}_u^{\dagger}\hat{a}_q\ket{\psi_0}\) as an example. Combining all possible contractions of a given type along with the fermionic parity signs (not shown here) leads to the determinant expressions in equations \ref{eq:fb_eff} and \ref{eq:eloc_wick}. The same types of terms arise for higher than doubly excited configurations and the adjacent column shows the optimal cost scaling for calculating each type for a general CI trial with \(N_c\) configurations. }\label{tab:scaling}
\centering
\begin{tabular}{ccccccc}
\hline
%Quantity && SMW && This work &~~& Tensor diagrams for doubly excited configurations\\
Quantity && SMW && \multicolumn{3}{c}{This work}\\
\cline{5-7}
& & & & Scaling &~~& Tensor diagrams for doubly excited configurations \\
\hline
\multicolumn{7}{c}{}\\
\multirow[t]{2}{*}{Force bias} && \multirow[t]{2}{*}{\(NM(N_c + X)\)} && \(N_c + XNM\) && \raisebox{-.3\height}{\includegraphics[width=0.52\textwidth]{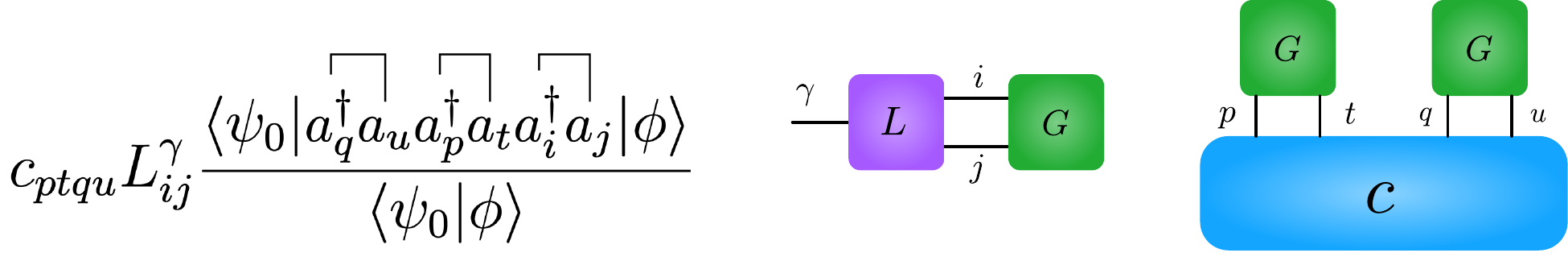}} \\
&& && \(N_c + XM^2\) && \raisebox{-.3\height}{\includegraphics[width=0.52\textwidth]{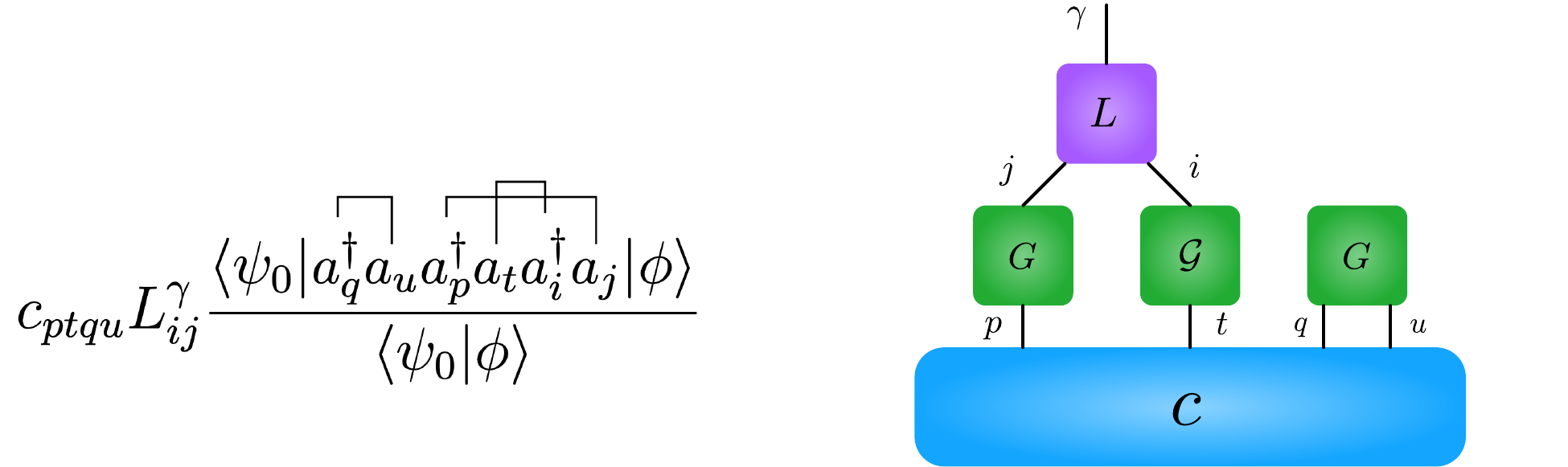}}\\
\multicolumn{7}{c}{}\\
\hline
\multicolumn{7}{c}{}\\
\multirow[t]{5}{*}{Local energy} && \multirow[t]{5}{*}{\(N_cXN^2M\)} && \(N_c + XNM\) && \raisebox{-.3\height}{\includegraphics[width=0.56\textwidth]{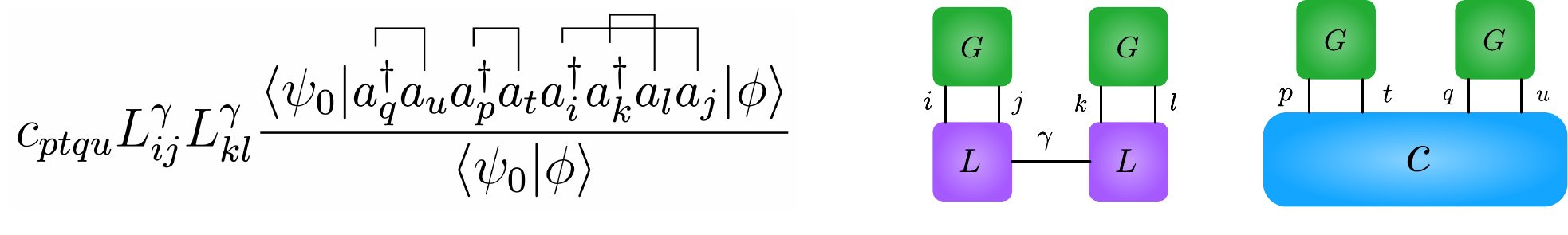}} \\
&& && \(N_c + XN^2M\) && \raisebox{-.4\height}{\includegraphics[width=0.56\textwidth]{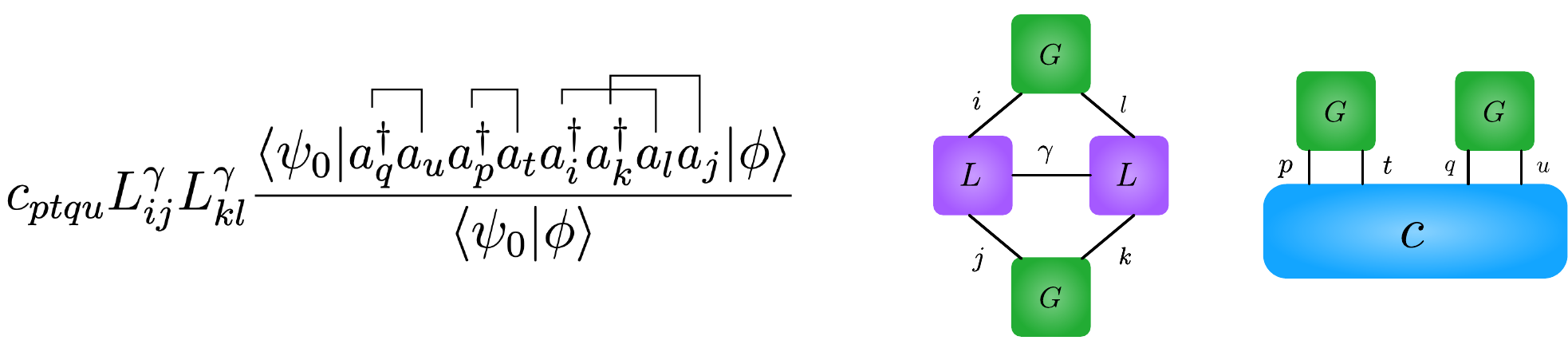}}\\
&& && \(N_c + XM^2\) && \raisebox{-.3\height}{\includegraphics[width=0.56\textwidth]{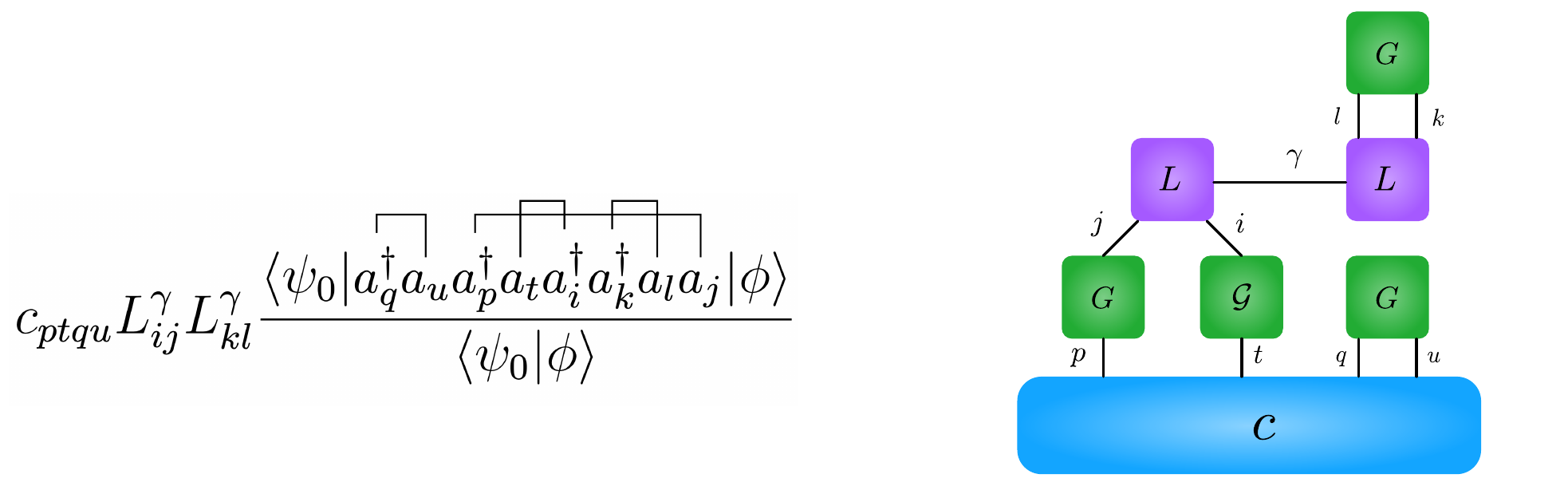}}\\
&& && \(N_c + XNM^2\) && \raisebox{-.3\height}{\includegraphics[width=0.56\textwidth]{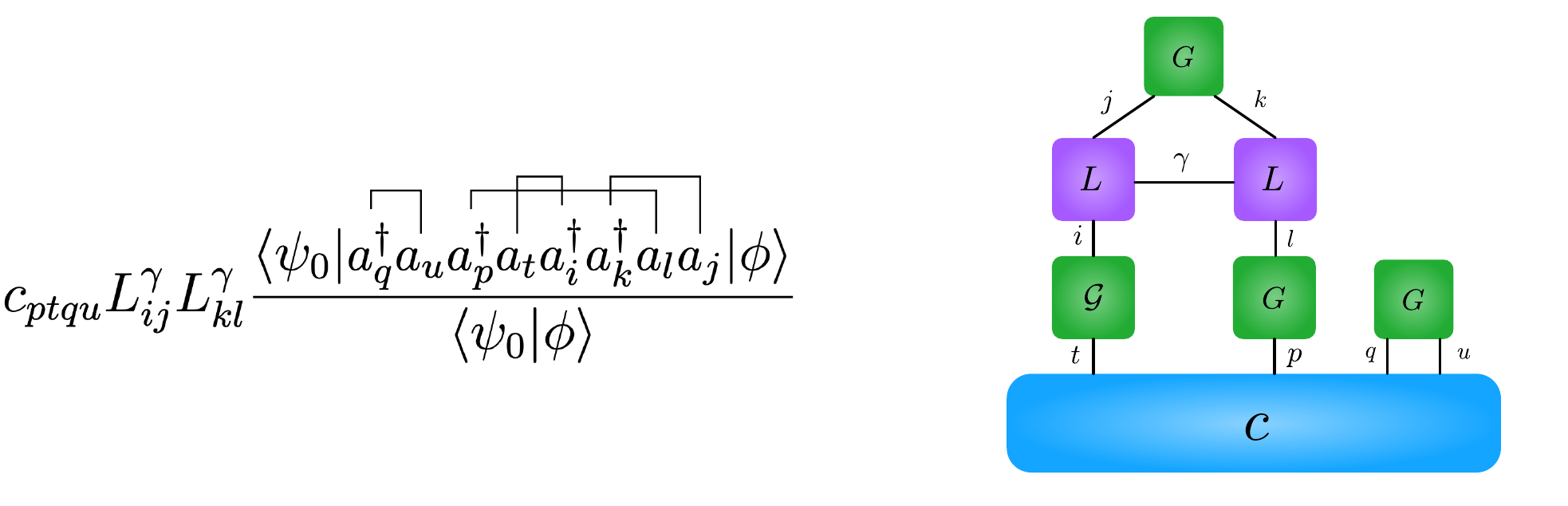}}\\
&& && \(X(N_c + NM^2)\) && \raisebox{-.3\height}{\includegraphics[width=0.56\textwidth]{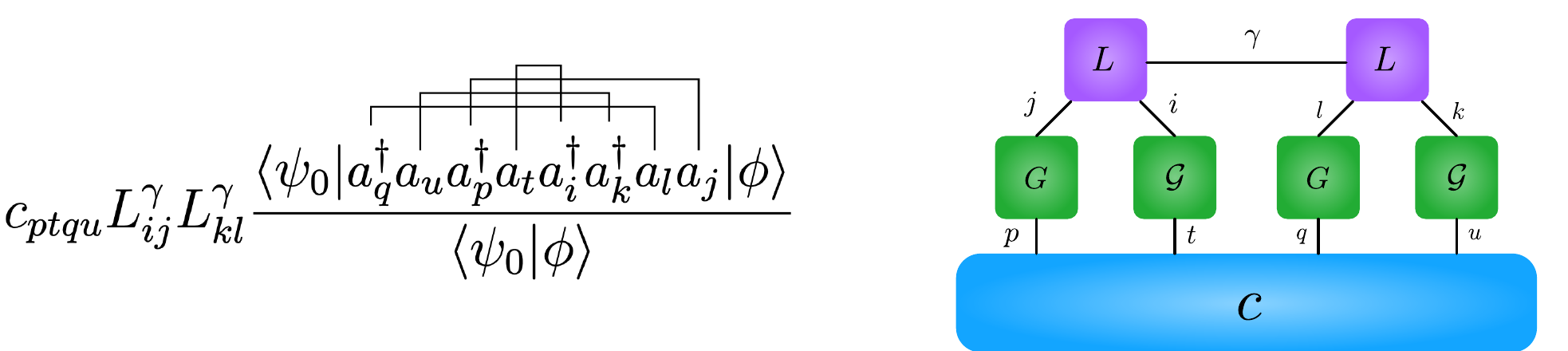}}\\
\multicolumn{7}{c}{}\\
\hline
\hline
\multicolumn{7}{l}{\textit{Key for notation}}\\
\multicolumn{7}{l}{\(N_c\): number of configurations, \(N\): number of electrons, \(M\): number of orbitals, \(X\): number of Cholesky vectors}\\
\multicolumn{7}{l}{\(c\): CI coefficients, \(L\): Cholesky vectors, \(G\) and \(\mathcal{G}\): Green's functions}\\
\multicolumn{7}{l}{\(i,j,k,l\): Hamiltonian indices; \(p,q,t,u\): CI excitation indices}\\
\hline
\end{tabular}
\end{table*}

\subsection{Local energy}\label{sec:eloc}

The calculation of local energy was discussed in detail in reference \onlinecite{mahajan2021taming}. Here, we give a brief summary of the algorithm. Consider the two-body part of the local energy (see equation \ref{eq:eloc}), expressed as
\begin{equation}
   E^2_L \left[\phi\right] = \frac{\sum_n\sum_{\gamma ijkl}c_nL^{\gamma}_{ij}L^{\gamma}_{kl}\frac{\expecth{\psi_n}{\hat{a}_i^{\dagger}\hat{a}_k^{\dagger}\hat{a}_l\hat{a}_j}{\phi}}{\inner{\psi_0}{\phi}}}{\sum_nc_n\frac{\inner{\psi_n}{\phi}}{\inner{\psi_0}{\phi}}}.
\end{equation}
The calculation of the overlap ratios in the denominator is described in the last section for the force bias. The matrix element in the numerator for the \(n\)th configuration is given by using the Wick's theorem as
\begin{equation}\label{eq:eloc_wick}
   \frac{\expecth{\psi_0}{\left(\prod_{\mu}^k \hat{a}^{\dagger}_{p_{\mu}}\hat{a}_{t_{\mu}}\right)\hat{a}_i^{\dagger}\hat{a}_k^{\dagger}\hat{a}_l\hat{a}_j}{\phi}}{\inner{\psi_0}{\phi}} = \det \begin{pmatrix}
		 G^{\left\{i,k\right\}}_{\left\{j,l\right\}} & \mathcal{G}^{\left\{i,k\right\}}_{\left\{t_{\mu}\right\}}\\[1em]
		 G^{\left\{p_{\mu}\right\}}_{\left\{j,l\right\}} & G^{\left\{p_{\mu}\right\}}_{\left\{t_{\mu}\right\}}\\
	 \end{pmatrix},
\end{equation}
where we have again dropped the configuration index subscript. Similar to the force bias calculation, the terms in this determinant can be split into two groups based on the kinds of pairwise contractions: those that do not involve cross contractions between CI and Hamiltonian excitation operators and those that involve at least one such cross contraction. The second group containing cross contractions can be further split into two groups depending on whether one or both of the Hamiltonian excitations are cross contracted with CI excitations. The resulting terms are shown as a tensor diagram in table \ref{tab:scaling}. Again, we have used all double excited CI configurations only for representational convenience. Different orders of performing the tensor contractions lead to different cost scalings, and using the one described in reference \onlinecite{mahajan2021taming} leads to a scaling of \(O(N_cX + XNM^2)\). For a selected CI state restricted to an active space of size \(A\), this cost is reduced to \(O(N_cX + XNAM + XN^2M)\).

\section{Results}\label{sec:results}
In this section, we present the results of our ph-AFQMC/HCI calculations and analyze the convergence of the phaseless error for several systems. In particular, we consider the utility of active space trial states for obtaining accurate energy differences and dipole moments. We used PySCF\cite{sun2018pyscf} to obtain molecular integrals and to perform all quantum chemistry wave function calculations. The SHCI code Dice was used to obtain the trial states. The code used to perform ph-AFQMC calculations is available in a public repository.\cite{dqmc_code} Input and output files for all calculations can also be accessed from a public repository.\cite{afqmc_files} Sources of systematic errors in ph-AFQMC calculations, besides the phaseless bias, include Trotter error, population control bias, errors in the Cholesky decomposition of the electron repulsion integrals, matrix exponential Taylor series truncation errors, and bias due to filtering of rare large fluctuations. We used a conservative time step of 0.005 a.u in all ph-AFQMC calculations. For population control, we used the reconfiguration procedure described in reference \onlinecite{buonaura1998numerical}. Cholesky decompositions were calculated up to a threshold error of \(10^{-5}\). Matrix exponentials were calculated by keeping terms up to the sixth power in the Taylor expansion. We filtered large fluctuations by capping weights and local energies.\cite{motta2018ab} Although it is difficult to get an accurate estimate of the systematic errors due to all these factors in general, they can be controlled systematically. We estimate these errors to be much smaller than the statistical errors in all the results presented here.

\subsection{Hydrogen chains}\label{sec:hchains}

\begin{figure}[htp]
\centering
\includegraphics[width=0.48\textwidth]{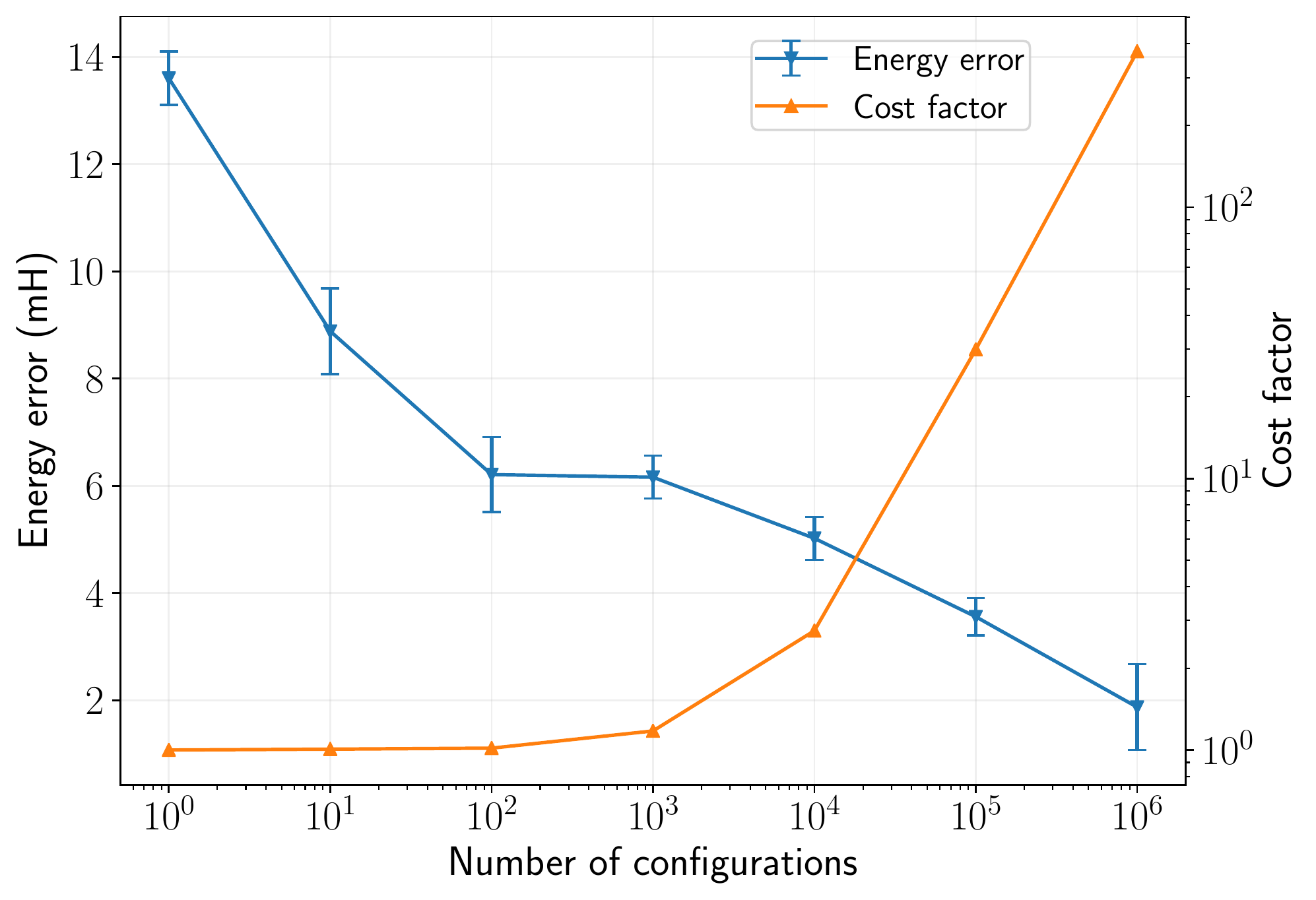}
\caption{Convergence of the ph-AFQMC/HCI phaseless energy error and scaling of the cost of computation with increasing number of configurations for \ce{H50} using the sto-6g basis at \(d=1.6\) Bohr. The phaseless error is calculated with respect to the near-exact DMRG energy, and the cost factor is calculated as the ratio of computational costs with respect to a single configuration calculation.}\label{fig:h50}
\end{figure}

Hydrogen chain systems are convenient for benchmarking due to their simplicity and availability of accurate density matrix renormalization group (DMRG) ground state energies.\cite{motta2017towards} First, we consider a chain of 50 equidistant Hydrogen atoms with an interatomic distance \(d = 1.6\) Bohr in the minimal sto-6g basis. We performed ph-AFQMC/HCI calculations on this system using progressively larger HCI expansion. These trial states were constructed as truncations of the HCI state obtained with \(\epsilon_1 = 10^{-4}\) in the canonical RHF basis. Figure \ref{fig:h50} shows the convergence of the phaseless error (calculated as a difference to the DMRG energy reported in reference ) with the number of configurations in the trial state. We see a nearly monotonic decrease in the phaseless error and the error is less than 2 mH for the trial with \(10^6\) configurations. The figure also shows the scaling of the cost of the ph-AFQMC calculation against the number of configurations with respect to the cost of a single configuration calculation. Remarkably, the cost factor for up to \(10^4\) configurations is less than 10 and increases roughly linearly with the number of configurations thereafter. This observation is consistent with the scaling relations discussed in section \ref{sec:sci_trial}, with the \(O(N_cX)\) factor in the cost scaling of local energy evaluation dominating for large HCI expansions. We found changing the value of \(\epsilon_1\) or slightly changing the selection criterion for choosing the configurations to only make a difference in ph-AFQMC energies for smaller number of configurations, leading to similar convergence behaviors for longer expansions.

\begin{figure}[htp]
\centering
\includegraphics[width=0.48\textwidth]{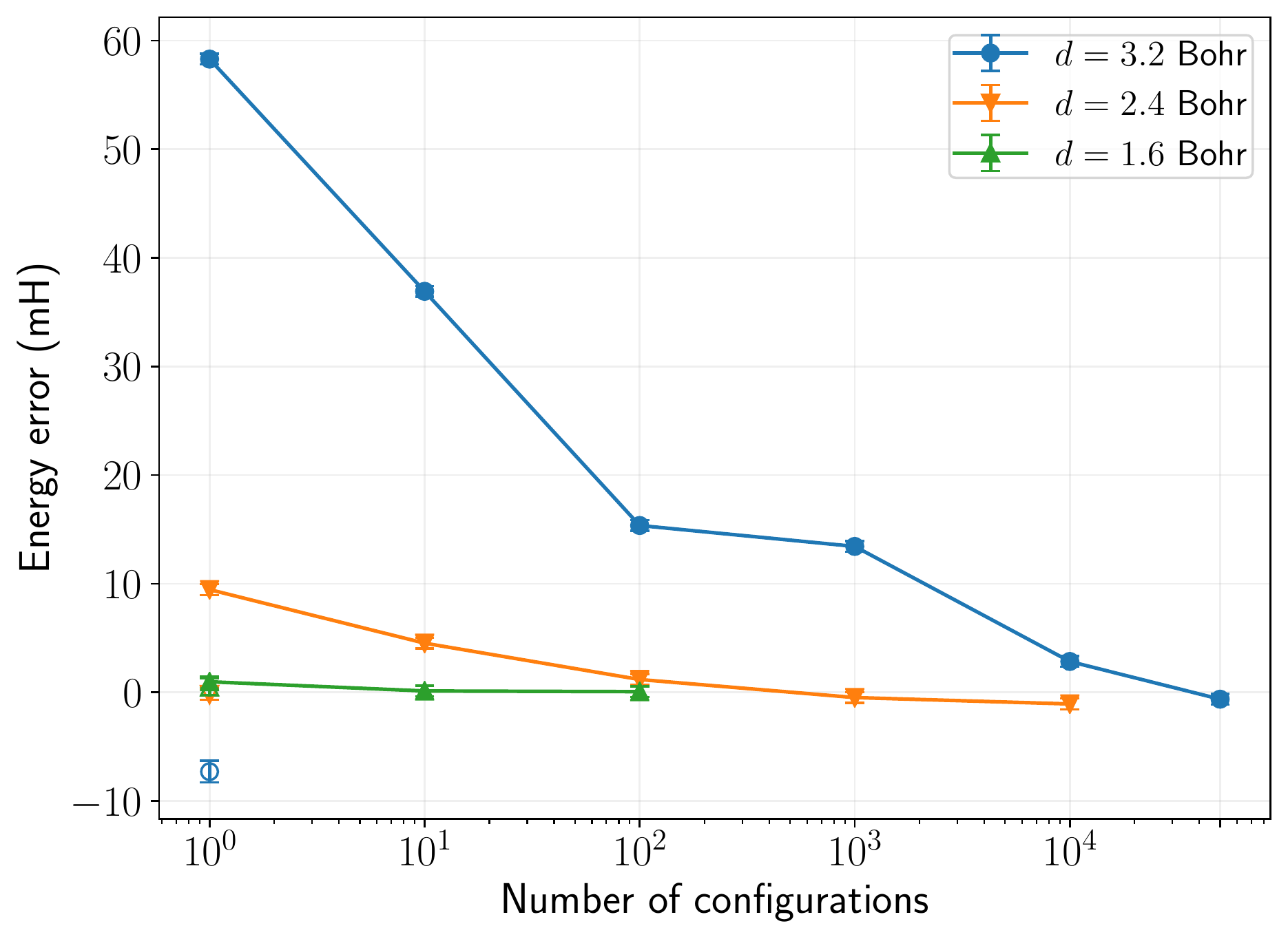}
\caption{ph-AFQMC/HCI phaseless errors for \ce{H10} using the cc-pVDZ basis set at different bond lengths. The trial states with different number of configurations are constructed by truncating a (10e, 10o) CASSCF wave function in each case. The hollow symbols show ph-AFQMC/UHF phaseless errors. (They are almost zero for \(d = 1.6\) Bohr and \(d = 2.4\) Bohr.)}\label{fig:h10}
\end{figure}

We also studied the \ce{H10} chain in the cc-pVDZ basis at different bond lengths to gauge the performance of active space trial states. We first performed a (10e, 10o) CASSCF (complete active space self consistent field) calculation and generated trial states by truncating the active space wave function. Figure \ref{fig:h10} shows the convergence of the phaseless error with the size of the HCI trial. For all three bond lengths, we see a systematic convergence to an almost vanishing error with the active space trial. The shortest bond length near equilibrium converges rapidly, while the stretched bond length requires almost the full active space wave function to reach convergence. The out of active space phaseless error, arising due to not correlating any orbitals outside the active space in the trial state, is tiny in this case and, more importantly, is very similar for all three bond lengths. We also show ph-AFQMC/UHF phaseless errors in the figure for comparison. ph-AFQMC/UHF is known to be very accurate for hydrogen chains except at stretched geometries. At stretched geometries, a CI expansion based on an unrestricted HF reference has been employed in the past as a trial state\cite{motta2017towards} and could be a handy tool in other problems as well. The algorithms described in this work can be straightforwardly extended to work with UHF-based CI expansions.

\subsection{Transition metal oxides}\label{sec:tmo}
A recent study of transition metal oxide molecules reported benchmark ground state energies using various accurate many-body methods.\cite{williams2020direct} Here, we use ph-AFQMC/HCI to study convergence of the phaseless errors for active space trials in TiO, CrO, MnO, and FeO. The Trails-Needs pseudopotential and the corresponding dz basis sets\cite{trail2017shape} were used at equilibrium geometries used in reference \onlinecite{williams2020direct}.

\begin{figure}[htp]
\centering
\includegraphics[width=0.48\textwidth]{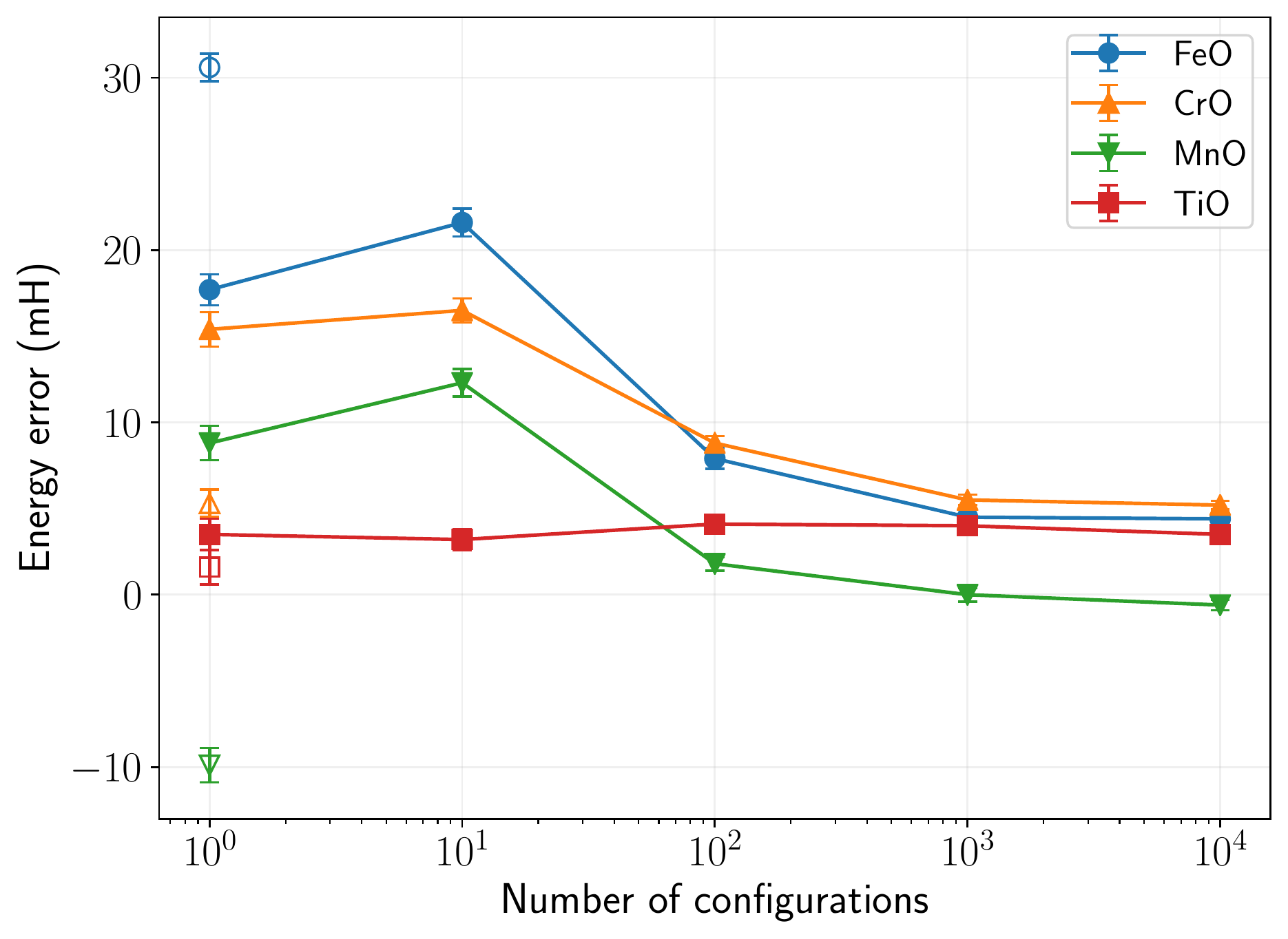}
\caption{Convergence of ph-AFQMC/HCI phaseless energy errors for four transition metal oxide molecules using the Trail-Needs dz pseudopotential and basis set. SHCI reference energies were used as references for calculating the phaseless errors. The trial states were constructed by truncating active space HCI wave functions. Hollow symbols show ph-AFQMC/UHF errors.}\label{fig:tmo}
\end{figure}

Figure \ref{fig:tmo} shows the convergence of phaseless errors with the number of configurations in the trial state at equilibrium geometries used in reference \onlinecite{williams2020direct}. We first performed CASSCF calculations with active spaces consisting of the metal \(3d\) and \(4d\) orbitals and oxygen \(2p\) orbitals. The HCI trial states were then generated in a space including the CASSCF active orbitals and the core orbitals, including all electrons. \(\epsilon_1 = 10^{-5}\) was used for the HCI calculations. Interestingly, for FeO, CrO, and MnO the phaseless error first increases before converging to a fixed value as the number of configurations in the trial is increased. This peculiar behavior highlights the fact that a lower energy trial state does not guarantee a smaller phaseless error. We note that the trends for smaller expansions are dependent on the particular schemes used for choosing the orbital spaces and for truncating the HCI states. But results for a large number of configurations are largely independent of such somewhat subjective choices. For TiO, using more configurations makes little difference to the phaseless error. We also performed ph-AFQMC/UHF calculations for comparison and found the phaseless errors to be 2(1), 5.3(8), -10(1), and 30.6(9) mH for TiO, CrO, MnO, and FeO, respectively.

The residual phaseless errors can be attributed solely to the out of HCI active space correlation. These errors can be systematically reduced by increasing the size of the active space, but getting vanishing errors this way is unlikely to be practical for larger systems. On the other hand, as noted in section \ref{sec:hchains}, it is reasonable to expect near-cancellation of this out of active space error when energy differences at different geometries are calculated using the same active space. Similar trends have been observed in multireference perturbation theories. To gauge the efficacy of this heuristic, we calculated ph-AFQMC/HCI energies of CrO at three bond lengths using the procedure described above. Table \ref{tab:cro} shows that the converged phaseless errors are almost identical at the three points. We note that it is crucial to converge the phaseless error of the active space trial, and different geometries may require trial states with different numbers of configurations to achieve convergence. This evidence suggests that in large problems, scalability issues of HCI trial states can be mitigated using active spaces.

\begin{table}
\caption{\ce{CrO} ground state energies and out of active space phaseless errors at different bond lengths. Bond lengths (\(d\)) are in Bohr, ph-AFQMC/HCI and SHCI energies are in H, the phaseless errors are in mH.}\label{tab:cro}
\centering
\begin{tabular}{ccccccc}
\hline
\(d\) &~~& ph-AFQMC/HCI &~~& SHCI &~~& Phaseless error \\
\hline
2.65 && -102.4993(5) && -102.5040(5) && 4.6(7) \\
3.06 && -102.5531(5) && -102.5584\cite{williams2020direct} && 5.3(5) \\
3.59 && -102.5269(7) && -102.5325(5) && 5.6(8) \\
\hline
\end{tabular}
\end{table}

\subsection{Dipole moments}\label{sec:dipoles}
The mixed estimator of an observable \(\hat{O}\) in the ground state is given by
\begin{equation}
   \langle O\rangle_{\text{mixed}} = \frac{\expecth{\psi_T}{\hat{O}}{\Psi_0}}{\inner{\psi_T}{\Psi_0}},
\end{equation}
where \(\ket{\Psi_0}\) is the ground state. In ph-AFQMC, one has access to an approximate sampling of the ground state, making the calculation of this mixed estimator convenient:

\begin{equation}
  \langle O\rangle_{\text{mixed}} \approx \frac{\sum_i w_i O_L(\phi_i)}{\sum_i w_i},
\end{equation}
where
\begin{equation}
   O_L(\phi_i) = \frac{\expecth{\psi_T}{\hat{O}}{\phi_i}}{\inner{\psi_T}{\phi_i}}
\end{equation}
is the local observable value.

For observables \(\hat{O}\) that do not commute with the system Hamiltonian, the mixed estimator has a systematic bias due to the trial state used to measure the observable. This bias can be reduced by using the extrapolated estimator, often employed in diffusion Monte Carlo,\cite{whitlock1979properties,ceperley1986quantum,rothstein2013survey} given by
\begin{equation}
   \langle \hat{O}\rangle_{\text{extrapolated}} = 2 \frac{\expecth{\psi_T}{\hat{O}}{\Psi_0}}{\inner{\psi_T}{\Psi_0}} - \frac{\expecth{\psi_T}{\hat{O}}{\psi_T}}{\inner{\psi_T}{\psi_T}},
\end{equation}
where the variational estimator of the trial state is used to correct the mixed estimator. The accuracy of this extrapolation depends critically on the quality of the trial state. Since ph-AFQMC calculations on molecular systems are typically performed using crude single determinant trials, extrapolation is not a viable option, and backpropagation is used to evaluate variational estimators of the sampled state instead.\cite{motta2017computation} Here, we study the behavior of the mixed and extrapolated estimators of dipole moments of a few small molecules. Evaluation of the local dipole moment for an HCI trial state is similar to the force bias calculation outlined in section \ref{sec:fb} since they both involve matrix elements of one-electron operators.

\begin{figure}[htp]
\centering
\includegraphics[width=0.48\textwidth]{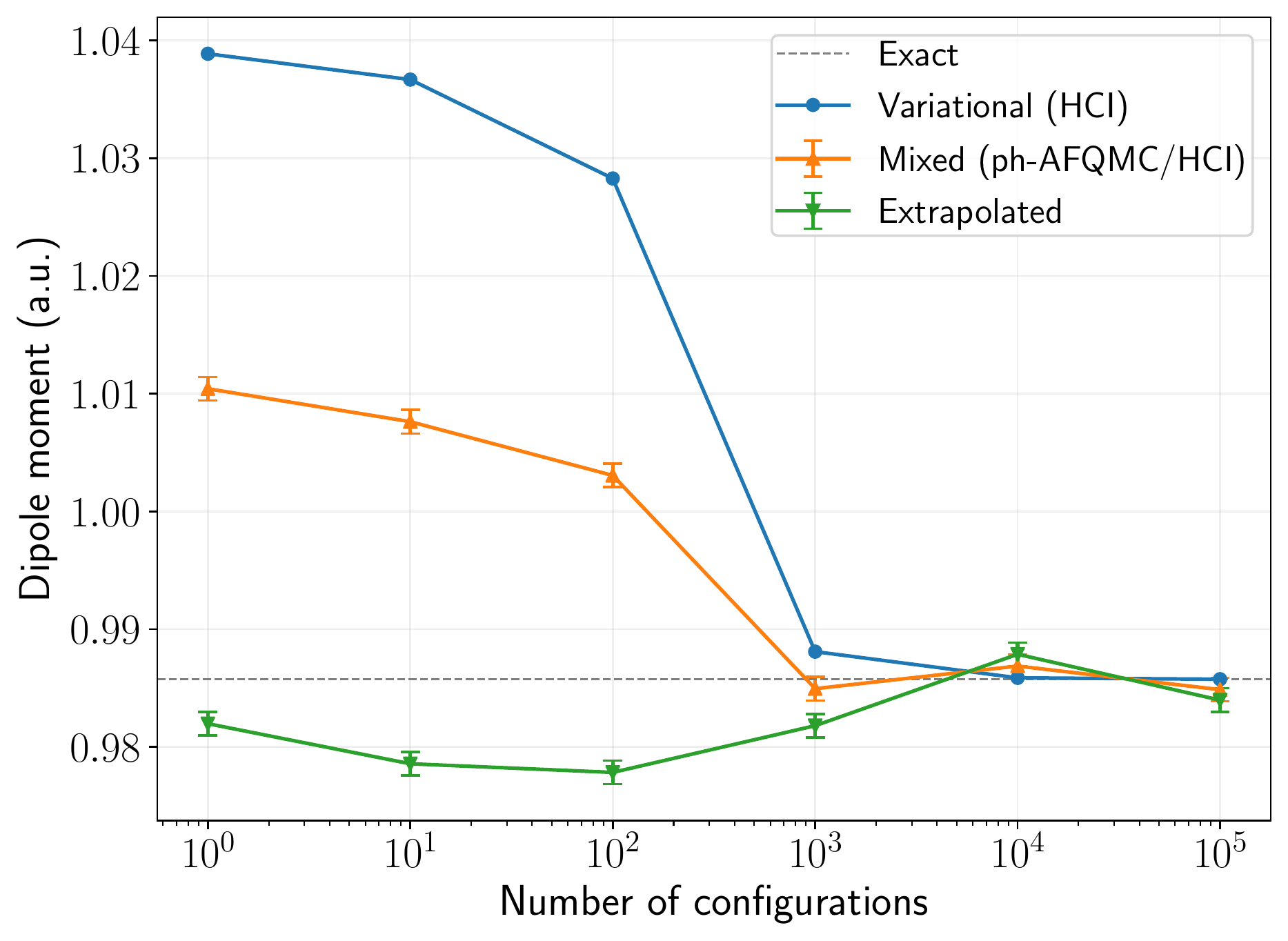}
\caption{Convergence of different estimators for the dipole moment of \ce{NH3} using the 6-31g basis. Variational and mixed estimators are calculated using HCI and ph-AFQMC/HCI, respectively.}\label{fig:nh3}
\end{figure}

We first look at the exactly solvable problem of \ce{NH3} in the 6-31g basis (geometry in the SI). We fully diagonalize this (10e, 14o) problem and calculate its ground state dipole moment. Mixed and extrapolated ph-AFQMC estimators were evaluated with trial states consisting of an increasing number of leading configurations from the ground state in the RHF canonical orbital basis. Figure \ref{fig:nh3} shows the convergence of the estimators with the trial state. In this case, we see a systematic convergence of the variational and mixed estimators to the exact value. The extrapolation works reasonably well, with consistently smaller errors than the mixed estimator.

\begin{table*}
\caption{Dipole moments (in a.u.) of three molecules using the aug-cc-pVQZ basis set. The reported ph-AFQMC/HCI values are converged with the number of configurations in the 50 orbital active space trial state.}\label{tab:dipoles}
\centering
\begin{tabular}{*{14}c}
\hline
Molecule &~~& \multicolumn{2}{c}{ph-AFQMC/HCI} &~~& RHF &~~& MP2 &~~& CCSD &~~& CCSD(T) &~~& Experiment \\
\cline{3-4}
& & Mixed & Extrapolated & & & & & & & & & & \\
\hline
\ce{CO} && 0.070(3) & 0.044(3) && -0.104 && 0.108 && 0.024 && 0.048 && 0.048\cite{muenter1975electric} \\
\ce{BF} && 0.356(4) & 0.321(4) && 0.333 && 0.377 && 0.314 && 0.325 && - \\
\ce{H2O} && 0.726(2) & 0.734(2) && 0.780 && 0.733 && 0.738 && 0.729 && 0.730\cite{lide2004crc} \\
\hline
\end{tabular}
\end{table*}

\begin{figure}[htp]
\centering
\includegraphics[width=0.48\textwidth]{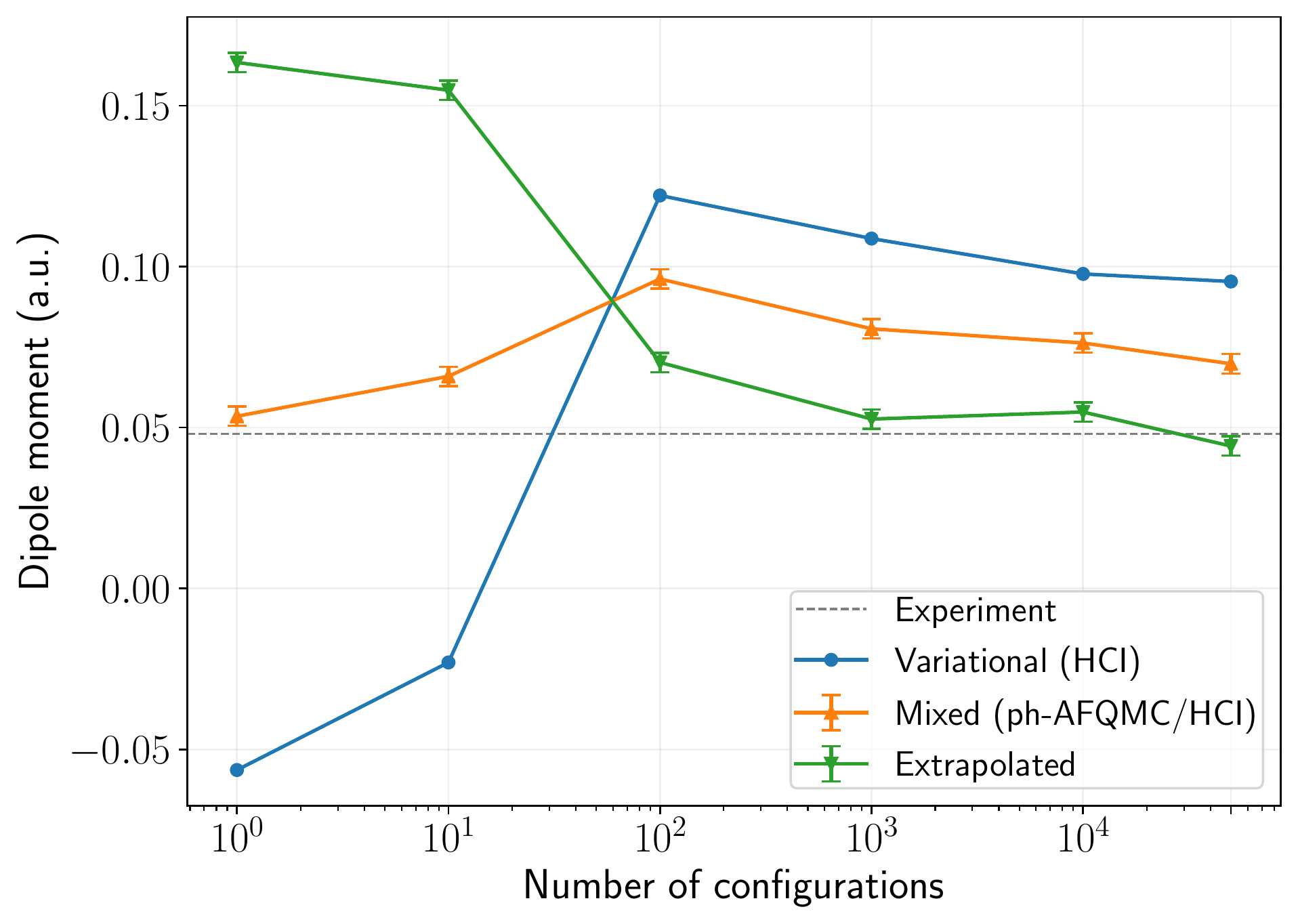}
\caption{Convergence of different estimators for the dipole moment of \ce{CO} using the aug-cc-pVQZ basis. Variational and mixed estimators are calculated using HCI and ph-AFQMC/HCI, respectively. Trial states are obtained by truncating a (14e, 50o) active space HCI wave function.}\label{fig:co}
\end{figure}

Table \ref{tab:dipoles} shows the calculated dipole moments for three molecules at equilibrium geometries in the aug-cc-pVQZ basis. The large basis set precludes exact evaluation, but CCSD(T) moments are known to be very accurate for the molecules considered here and agree well with the experimental values. We computed orbital relaxed coupled cluster and MP2 dipole moments by calculating energies at two different electric fields and evaluating the numerical derivative. For obtaining ph-AFQMC/HCI estimators, we first performed a full-valence CASSCF calculations. The trial states were then constructed using HCI to correlate all electrons in a space of 50 CASSCF orbitals. In all cases, both mixed and extrapolated estimators were nearly converged with the number of configurations in the active space (details can be found in the SI). Consider the case of the challenging \ce{CO} molecule, which has a small dipole moment, and RHF predicts the wrong direction polarity in this case. The convergence of the dipole moment estimators with number of configurations in the trial state is shown in figure \ref{fig:co}. For a small number of configurations (fewer than 100), the variational estimator is qualitatively incorrect similar to RHF. Adding more configurations flips the sign, and the variational estimator seems to eventually converge to the asymptotic value for the truncated space of 50 orbitals. The mixed estimator also exhibits a non-monotonic convergence with its value for a single determinant close to the experimental value because of a fortuitous cancellation of mixed estimator and phaseless biases. The extrapolated estimator for small number of configurations in the trial is very poor due to the erroneous variational estimates. For more than \(10^3\) configurations, while the mixed estimator seems to be converging to a substantially biased dipole moment value, the extrapolation rectifies this bias reasonably well. The extrapolated estimators are in good agreement with the CCSD(T) and experimental values for all three molecules. Evidently, extrapolation effectively corrects the out of active space bias in the mixed estimator for \ce{CO} and \ce{BF}.

\section{Conclusion}\label{sec:conclusion}
In this work, we presented efficient algorithms for using selected CI trial states in ph-AFQMC. We demonstrated their favorable scaling by showing that simulations with long sCI expansions incur little overhead. In our analysis of the convergence of phaseless error in ground-state energies and dipole moments as a function of the number of configurations, we found it to be non-monotonic in some cases, highlighting the importance of using a sufficient number of determinants in these systems. The use of states restricted to moderately sized active spaces yielded excellent results for energy differences and dipole moments. Our numerical experiments suggest that this may be a practical way to tackle larger systems, where the generation and handling of sCI states are challenging. Finally, we showed that the extrapolated estimator for dipole moments is accurate if enough configurations are present in the trial state. If this behavior turns out to be true in general, extrapolated estimators could be a viable alternative to backpropagation for calculating ground state properties.

Our technique can be employed in different ways for studying challenging systems. It can be used to validate the use of simpler and less expensive trial states in ph-AFQMC for large calculations. ph-AFQMC/HCI can be adopted as an accurate solver in embedding schemes for describing the embedded cluster. It would be interesting to study the feasibility of the extrapolated estimator in larger systems and for calculating two-body properties. Lastly, we note that this technique can be utilized in the quantum-classical hybrid AFQMC (QC-AFQMC) in prototypical and practical quantum computations.\cite{Huggins2021Jun}

\section*{Data availability}
The data that support the findings of this study are available in a public repository at \onlinecite{afqmc_files}.

\section*{Acknowledgements}
AM and SS were supported by the National Science Foundation through the grant CHE-1800584. SS was also partly supported through the Sloan research fellowship. JL thanks David Reichman for support and encouragement. All calculations were performed on the Blanca and Summit clusters at CU Boulder.

%\bibliographystyle{achemso}
%\bibliography{ref}

\begin{mcitethebibliography}{56}
\providecommand*\natexlab[1]{#1}
\providecommand*\mciteSetBstSublistMode[1]{}
\providecommand*\mciteSetBstMaxWidthForm[2]{}
\providecommand*\mciteBstWouldAddEndPuncttrue
  {\def\EndOfBibitem{\unskip.}}
\providecommand*\mciteBstWouldAddEndPunctfalse
  {\let\EndOfBibitem\relax}
\providecommand*\mciteSetBstMidEndSepPunct[3]{}
\providecommand*\mciteSetBstSublistLabelBeginEnd[3]{}
\providecommand*\EndOfBibitem{}
\mciteSetBstSublistMode{f}
\mciteSetBstMaxWidthForm{subitem}{(\alph{mcitesubitemcount})}
\mciteSetBstSublistLabelBeginEnd
  {\mcitemaxwidthsubitemform\space}
  {\relax}
  {\relax}

\bibitem[Kalos \latin{et~al.}(1974)Kalos, Levesque, and
  Verlet]{kalos1974helium}
Kalos,~M.~H.; Levesque,~D.; Verlet,~L. Helium at zero temperature with
  hard-sphere and other forces. \emph{Physical Review A} \textbf{1974},
  \emph{9}, 2178\relax
\mciteBstWouldAddEndPuncttrue
\mciteSetBstMidEndSepPunct{\mcitedefaultmidpunct}
{\mcitedefaultendpunct}{\mcitedefaultseppunct}\relax
\EndOfBibitem
\bibitem[Ceperley \latin{et~al.}(1977)Ceperley, Chester, and
  Kalos]{ceperley1977monte}
Ceperley,~D.; Chester,~G.~V.; Kalos,~M.~H. Monte Carlo simulation of a
  many-fermion study. \emph{Physical Review B} \textbf{1977}, \emph{16},
  3081\relax
\mciteBstWouldAddEndPuncttrue
\mciteSetBstMidEndSepPunct{\mcitedefaultmidpunct}
{\mcitedefaultendpunct}{\mcitedefaultseppunct}\relax
\EndOfBibitem
\bibitem[Ceperley and Alder(1980)Ceperley, and Alder]{ceperley1980ground}
Ceperley,~D.~M.; Alder,~B.~J. Ground state of the electron gas by a stochastic
  method. \emph{Physical review letters} \textbf{1980}, \emph{45}, 566\relax
\mciteBstWouldAddEndPuncttrue
\mciteSetBstMidEndSepPunct{\mcitedefaultmidpunct}
{\mcitedefaultendpunct}{\mcitedefaultseppunct}\relax
\EndOfBibitem
\bibitem[Nightingale and Umrigar(1998)Nightingale, and
  Umrigar]{nightingale1998quantum}
Nightingale,~M.~P.; Umrigar,~C.~J. \emph{Quantum Monte Carlo methods in physics
  and chemistry}; Springer Science \& Business Media, 1998\relax
\mciteBstWouldAddEndPuncttrue
\mciteSetBstMidEndSepPunct{\mcitedefaultmidpunct}
{\mcitedefaultendpunct}{\mcitedefaultseppunct}\relax
\EndOfBibitem
\bibitem[Foulkes \latin{et~al.}(2001)Foulkes, Mitas, Needs, and
  Rajagopal]{foulkes2001quantum}
Foulkes,~W.; Mitas,~L.; Needs,~R.; Rajagopal,~G. Quantum Monte Carlo
  simulations of solids. \emph{Reviews of Modern Physics} \textbf{2001},
  \emph{73}, 33\relax
\mciteBstWouldAddEndPuncttrue
\mciteSetBstMidEndSepPunct{\mcitedefaultmidpunct}
{\mcitedefaultendpunct}{\mcitedefaultseppunct}\relax
\EndOfBibitem
\bibitem[Booth \latin{et~al.}(2013)Booth, Gr{\"u}neis, Kresse, and
  Alavi]{booth2013towards}
Booth,~G.~H.; Gr{\"u}neis,~A.; Kresse,~G.; Alavi,~A. Towards an exact
  description of electronic wavefunctions in real solids. \emph{Nature}
  \textbf{2013}, \emph{493}, 365\relax
\mciteBstWouldAddEndPuncttrue
\mciteSetBstMidEndSepPunct{\mcitedefaultmidpunct}
{\mcitedefaultendpunct}{\mcitedefaultseppunct}\relax
\EndOfBibitem
\bibitem[Becca and Sorella(2017)Becca, and Sorella]{becca2017quantum}
Becca,~F.; Sorella,~S. \emph{Quantum Monte Carlo approaches for correlated
  systems}; Cambridge University Press, 2017\relax
\mciteBstWouldAddEndPuncttrue
\mciteSetBstMidEndSepPunct{\mcitedefaultmidpunct}
{\mcitedefaultendpunct}{\mcitedefaultseppunct}\relax
\EndOfBibitem
\bibitem[Zhang and Krakauer(2003)Zhang, and Krakauer]{zhang2003quantum}
Zhang,~S.; Krakauer,~H. Quantum Monte Carlo method using phase-free random
  walks with Slater determinants. \emph{Physical review letters} \textbf{2003},
  \emph{90}, 136401\relax
\mciteBstWouldAddEndPuncttrue
\mciteSetBstMidEndSepPunct{\mcitedefaultmidpunct}
{\mcitedefaultendpunct}{\mcitedefaultseppunct}\relax
\EndOfBibitem
\bibitem[Fahy and Hamann(1990)Fahy, and Hamann]{fahy1990positive}
Fahy,~S.; Hamann,~D. Positive-projection Monte Carlo simulation: A new
  variational approach to strongly interacting fermion systems. \emph{Physical
  review letters} \textbf{1990}, \emph{65}, 3437\relax
\mciteBstWouldAddEndPuncttrue
\mciteSetBstMidEndSepPunct{\mcitedefaultmidpunct}
{\mcitedefaultendpunct}{\mcitedefaultseppunct}\relax
\EndOfBibitem
\bibitem[Zhang \latin{et~al.}(1995)Zhang, Carlson, and
  Gubernatis]{zhang1995constrained}
Zhang,~S.; Carlson,~J.; Gubernatis,~J.~E. Constrained path quantum Monte Carlo
  method for fermion ground states. \emph{Physical review letters}
  \textbf{1995}, \emph{74}, 3652\relax
\mciteBstWouldAddEndPuncttrue
\mciteSetBstMidEndSepPunct{\mcitedefaultmidpunct}
{\mcitedefaultendpunct}{\mcitedefaultseppunct}\relax
\EndOfBibitem
\bibitem[Zhang \latin{et~al.}(1997)Zhang, Carlson, and
  Gubernatis]{zhang1997constrained}
Zhang,~S.; Carlson,~J.; Gubernatis,~J.~E. Constrained path Monte Carlo method
  for fermion ground states. \emph{Physical Review B} \textbf{1997}, \emph{55},
  7464\relax
\mciteBstWouldAddEndPuncttrue
\mciteSetBstMidEndSepPunct{\mcitedefaultmidpunct}
{\mcitedefaultendpunct}{\mcitedefaultseppunct}\relax
\EndOfBibitem
\bibitem[Al-Saidi \latin{et~al.}(2006)Al-Saidi, Zhang, and
  Krakauer]{al2006auxiliary}
Al-Saidi,~W.; Zhang,~S.; Krakauer,~H. Auxiliary-field quantum Monte Carlo
  calculations of molecular systems with a Gaussian basis. \emph{The Journal of
  chemical physics} \textbf{2006}, \emph{124}, 224101\relax
\mciteBstWouldAddEndPuncttrue
\mciteSetBstMidEndSepPunct{\mcitedefaultmidpunct}
{\mcitedefaultendpunct}{\mcitedefaultseppunct}\relax
\EndOfBibitem
\bibitem[Al-Saidi \latin{et~al.}(2007)Al-Saidi, Zhang, and
  Krakauer]{al2007bond}
Al-Saidi,~W.~A.; Zhang,~S.; Krakauer,~H. Bond breaking with auxiliary-field
  quantum Monte Carlo. \emph{The Journal of chemical physics} \textbf{2007},
  \emph{127}, 144101\relax
\mciteBstWouldAddEndPuncttrue
\mciteSetBstMidEndSepPunct{\mcitedefaultmidpunct}
{\mcitedefaultendpunct}{\mcitedefaultseppunct}\relax
\EndOfBibitem
\bibitem[Suewattana \latin{et~al.}(2007)Suewattana, Purwanto, Zhang, Krakauer,
  and Walter]{suewattana2007phaseless}
Suewattana,~M.; Purwanto,~W.; Zhang,~S.; Krakauer,~H.; Walter,~E.~J. Phaseless
  auxiliary-field quantum Monte Carlo calculations with plane waves and
  pseudopotentials: Applications to atoms and molecules. \emph{Physical Review
  B} \textbf{2007}, \emph{75}, 245123\relax
\mciteBstWouldAddEndPuncttrue
\mciteSetBstMidEndSepPunct{\mcitedefaultmidpunct}
{\mcitedefaultendpunct}{\mcitedefaultseppunct}\relax
\EndOfBibitem
\bibitem[Purwanto \latin{et~al.}(2015)Purwanto, Zhang, and
  Krakauer]{purwanto2015auxiliary}
Purwanto,~W.; Zhang,~S.; Krakauer,~H. An auxiliary-field quantum Monte Carlo
  study of the chromium dimer. \emph{J. Chem. Phys.} \textbf{2015}, \emph{142},
  064302\relax
\mciteBstWouldAddEndPuncttrue
\mciteSetBstMidEndSepPunct{\mcitedefaultmidpunct}
{\mcitedefaultendpunct}{\mcitedefaultseppunct}\relax
\EndOfBibitem
\bibitem[Motta and Zhang(2018)Motta, and Zhang]{motta2018ab}
Motta,~M.; Zhang,~S. Ab initio computations of molecular systems by the
  auxiliary-field quantum Monte Carlo method. \emph{Wiley Interdisciplinary
  Reviews: Computational Molecular Science} \textbf{2018}, \emph{8},
  e1364\relax
\mciteBstWouldAddEndPuncttrue
\mciteSetBstMidEndSepPunct{\mcitedefaultmidpunct}
{\mcitedefaultendpunct}{\mcitedefaultseppunct}\relax
\EndOfBibitem
\bibitem[Hao \latin{et~al.}(2018)Hao, Shee, Upadhyay, Ataca, Jordan, and
  Rubenstein]{hao2018accurate}
Hao,~H.; Shee,~J.; Upadhyay,~S.; Ataca,~C.; Jordan,~K.~D.; Rubenstein,~B.~M.
  Accurate predictions of electron binding energies of dipole-bound anions via
  quantum Monte Carlo methods. \emph{The journal of physical chemistry letters}
  \textbf{2018}, \emph{9}, 6185--6190\relax
\mciteBstWouldAddEndPuncttrue
\mciteSetBstMidEndSepPunct{\mcitedefaultmidpunct}
{\mcitedefaultendpunct}{\mcitedefaultseppunct}\relax
\EndOfBibitem
\bibitem[Motta \latin{et~al.}(2019)Motta, Shee, Zhang, and
  Chan]{motta2019efficient}
Motta,~M.; Shee,~J.; Zhang,~S.; Chan,~G. K.-L. Efficient ab initio
  auxiliary-field quantum Monte Carlo calculations in Gaussian bases via
  low-rank tensor decomposition. \emph{Journal of chemical theory and
  computation} \textbf{2019}, \emph{15}, 3510--3521\relax
\mciteBstWouldAddEndPuncttrue
\mciteSetBstMidEndSepPunct{\mcitedefaultmidpunct}
{\mcitedefaultendpunct}{\mcitedefaultseppunct}\relax
\EndOfBibitem
\bibitem[Shee \latin{et~al.}(2019)Shee, Arthur, Zhang, Reichman, and
  Friesner]{shee2019singlet}
Shee,~J.; Arthur,~E.~J.; Zhang,~S.; Reichman,~D.~R.; Friesner,~R.~A.
  Singlet--triplet energy gaps of organic biradicals and polyacenes with
  auxiliary-field quantum Monte Carlo. \emph{Journal of chemical theory and
  computation} \textbf{2019}, \emph{15}, 4924--4932\relax
\mciteBstWouldAddEndPuncttrue
\mciteSetBstMidEndSepPunct{\mcitedefaultmidpunct}
{\mcitedefaultendpunct}{\mcitedefaultseppunct}\relax
\EndOfBibitem
\bibitem[Shee \latin{et~al.}(2019)Shee, Rudshteyn, Arthur, Zhang, Reichman, and
  Friesner]{shee2019achieving}
Shee,~J.; Rudshteyn,~B.; Arthur,~E.~J.; Zhang,~S.; Reichman,~D.~R.;
  Friesner,~R.~A. On Achieving High Accuracy in Quantum Chemical Calculations
  of 3 d Transition Metal-Containing Systems: A Comparison of Auxiliary-Field
  Quantum Monte Carlo with Coupled Cluster, Density Functional Theory, and
  Experiment for Diatomic Molecules. \emph{Journal of chemical theory and
  computation} \textbf{2019}, \emph{15}, 2346--2358\relax
\mciteBstWouldAddEndPuncttrue
\mciteSetBstMidEndSepPunct{\mcitedefaultmidpunct}
{\mcitedefaultendpunct}{\mcitedefaultseppunct}\relax
\EndOfBibitem
\bibitem[Williams \latin{et~al.}(2020)Williams, Yao, Li, Chen, Shi, Motta, Niu,
  Ray, Guo, Anderson, \latin{et~al.} others]{williams2020direct}
Williams,~K.~T.; Yao,~Y.; Li,~J.; Chen,~L.; Shi,~H.; Motta,~M.; Niu,~C.;
  Ray,~U.; Guo,~S.; Anderson,~R.~J., \latin{et~al.}  Direct comparison of
  many-body methods for realistic electronic Hamiltonians. \emph{Physical
  Review X} \textbf{2020}, \emph{10}, 011041\relax
\mciteBstWouldAddEndPuncttrue
\mciteSetBstMidEndSepPunct{\mcitedefaultmidpunct}
{\mcitedefaultendpunct}{\mcitedefaultseppunct}\relax
\EndOfBibitem
\bibitem[Lee and Reichman(2020)Lee, and Reichman]{lee2020stochastic}
Lee,~J.; Reichman,~D.~R. Stochastic resolution-of-the-identity auxiliary-field
  quantum Monte Carlo: Scaling reduction without overhead. \emph{The Journal of
  Chemical Physics} \textbf{2020}, \emph{153}, 044131\relax
\mciteBstWouldAddEndPuncttrue
\mciteSetBstMidEndSepPunct{\mcitedefaultmidpunct}
{\mcitedefaultendpunct}{\mcitedefaultseppunct}\relax
\EndOfBibitem
\bibitem[Lee \latin{et~al.}(2020)Lee, Malone, and Morales]{lee2020utilizing}
Lee,~J.; Malone,~F.~D.; Morales,~M.~A. Utilizing essential symmetry breaking in
  auxiliary-field quantum Monte Carlo: Application to the spin gaps of the C36
  fullerene and an iron porphyrin model complex. \emph{Journal of chemical
  theory and computation} \textbf{2020}, \emph{16}, 3019--3027\relax
\mciteBstWouldAddEndPuncttrue
\mciteSetBstMidEndSepPunct{\mcitedefaultmidpunct}
{\mcitedefaultendpunct}{\mcitedefaultseppunct}\relax
\EndOfBibitem
\bibitem[Shi and Zhang(2021)Shi, and Zhang]{shi2021some}
Shi,~H.; Zhang,~S. Some recent developments in auxiliary-field quantum Monte
  Carlo for real materials. \emph{The Journal of Chemical Physics}
  \textbf{2021}, \emph{154}, 024107\relax
\mciteBstWouldAddEndPuncttrue
\mciteSetBstMidEndSepPunct{\mcitedefaultmidpunct}
{\mcitedefaultendpunct}{\mcitedefaultseppunct}\relax
\EndOfBibitem
\bibitem[Loh~Jr \latin{et~al.}(1990)Loh~Jr, Gubernatis, Scalettar, White,
  Scalapino, and Sugar]{loh1990sign}
Loh~Jr,~E.; Gubernatis,~J.; Scalettar,~R.; White,~S.; Scalapino,~D.; Sugar,~R.
  Sign problem in the numerical simulation of many-electron systems.
  \emph{Physical Review B} \textbf{1990}, \emph{41}, 9301\relax
\mciteBstWouldAddEndPuncttrue
\mciteSetBstMidEndSepPunct{\mcitedefaultmidpunct}
{\mcitedefaultendpunct}{\mcitedefaultseppunct}\relax
\EndOfBibitem
\bibitem[Troyer and Wiese(2005)Troyer, and Wiese]{troyer2005computational}
Troyer,~M.; Wiese,~U.-J. Computational complexity and fundamental limitations
  to fermionic quantum Monte Carlo simulations. \emph{Physical review letters}
  \textbf{2005}, \emph{94}, 170201\relax
\mciteBstWouldAddEndPuncttrue
\mciteSetBstMidEndSepPunct{\mcitedefaultmidpunct}
{\mcitedefaultendpunct}{\mcitedefaultseppunct}\relax
\EndOfBibitem
\bibitem[Mahajan and Sharma(2021)Mahajan, and Sharma]{mahajan2021taming}
Mahajan,~A.; Sharma,~S. Taming the Sign Problem in Auxiliary-Field Quantum
  Monte Carlo Using Accurate Wave Functions. \emph{Journal of Chemical Theory
  and Computation} \textbf{2021}, \emph{17}, 4786--4798\relax
\mciteBstWouldAddEndPuncttrue
\mciteSetBstMidEndSepPunct{\mcitedefaultmidpunct}
{\mcitedefaultendpunct}{\mcitedefaultseppunct}\relax
\EndOfBibitem
\bibitem[Giner \latin{et~al.}(2013)Giner, Scemama, and
  Caffarel]{giner2013using}
Giner,~E.; Scemama,~A.; Caffarel,~M. Using perturbatively selected
  configuration interaction in quantum Monte Carlo calculations. \emph{Canadian
  Journal of Chemistry} \textbf{2013}, \emph{91}, 879--885\relax
\mciteBstWouldAddEndPuncttrue
\mciteSetBstMidEndSepPunct{\mcitedefaultmidpunct}
{\mcitedefaultendpunct}{\mcitedefaultseppunct}\relax
\EndOfBibitem
\bibitem[Evangelista(2014)]{evangelista2014adaptive}
Evangelista,~F.~A. Adaptive multiconfigurational wave functions. \emph{The
  Journal of Chemical Physics} \textbf{2014}, \emph{140}, 124114\relax
\mciteBstWouldAddEndPuncttrue
\mciteSetBstMidEndSepPunct{\mcitedefaultmidpunct}
{\mcitedefaultendpunct}{\mcitedefaultseppunct}\relax
\EndOfBibitem
\bibitem[Holmes \latin{et~al.}(2016)Holmes, Tubman, and Umrigar]{Holmes2016b}
Holmes,~A.~A.; Tubman,~N.~M.; Umrigar,~C.~J. Heat-Bath Configuration
  Interaction: An Efficient Selected Configuration Interaction Algorithm
  Inspired by Heat-Bath Sampling. \emph{J. Chem. Theory Comput.} \textbf{2016},
  \emph{12}, 3674--3680, PMID: 27428771\relax
\mciteBstWouldAddEndPuncttrue
\mciteSetBstMidEndSepPunct{\mcitedefaultmidpunct}
{\mcitedefaultendpunct}{\mcitedefaultseppunct}\relax
\EndOfBibitem
\bibitem[Tubman \latin{et~al.}(2016)Tubman, Lee, Takeshita, Head-Gordon, and
  Whaley]{tubman2016deterministic}
Tubman,~N.~M.; Lee,~J.; Takeshita,~T.~Y.; Head-Gordon,~M.; Whaley,~K.~B. A
  deterministic alternative to the full configuration interaction quantum Monte
  Carlo method. \emph{The Journal of chemical physics} \textbf{2016},
  \emph{145}, 044112\relax
\mciteBstWouldAddEndPuncttrue
\mciteSetBstMidEndSepPunct{\mcitedefaultmidpunct}
{\mcitedefaultendpunct}{\mcitedefaultseppunct}\relax
\EndOfBibitem
\bibitem[Filippi \latin{et~al.}(2016)Filippi, Assaraf, and
  Moroni]{filippi2016simple}
Filippi,~C.; Assaraf,~R.; Moroni,~S. Simple formalism for efficient derivatives
  and multi-determinant expansions in quantum Monte Carlo. \emph{The Journal of
  chemical physics} \textbf{2016}, \emph{144}, 194105\relax
\mciteBstWouldAddEndPuncttrue
\mciteSetBstMidEndSepPunct{\mcitedefaultmidpunct}
{\mcitedefaultendpunct}{\mcitedefaultseppunct}\relax
\EndOfBibitem
\bibitem[Assaraf \latin{et~al.}(2017)Assaraf, Moroni, and Filippi]{Assaraf2017}
Assaraf,~R.; Moroni,~S.; Filippi,~C. Optimizing the Energy with Quantum Monte
  Carlo: A Lower Numerical Scaling for Jastrow–Slater Expansions. \emph{J.
  Chem. Theory Comput.} \textbf{2017}, \emph{13}, 5273--5281, PMID:
  28873307\relax
\mciteBstWouldAddEndPuncttrue
\mciteSetBstMidEndSepPunct{\mcitedefaultmidpunct}
{\mcitedefaultendpunct}{\mcitedefaultseppunct}\relax
\EndOfBibitem
\bibitem[Dash \latin{et~al.}(2018)Dash, Moroni, Scemama, and
  Filippi]{dash2018perturbatively}
Dash,~M.; Moroni,~S.; Scemama,~A.; Filippi,~C. Perturbatively selected
  configuration-interaction wave functions for efficient geometry optimization
  in quantum Monte Carlo. \emph{Journal of chemical theory and computation}
  \textbf{2018}, \emph{14}, 4176--4182\relax
\mciteBstWouldAddEndPuncttrue
\mciteSetBstMidEndSepPunct{\mcitedefaultmidpunct}
{\mcitedefaultendpunct}{\mcitedefaultseppunct}\relax
\EndOfBibitem
\bibitem[Pineda~Flores and Neuscamman(2019)Pineda~Flores, and
  Neuscamman]{pineda2019excited}
Pineda~Flores,~S.~D.; Neuscamman,~E. Excited state specific multi-Slater
  Jastrow wave functions. \emph{The Journal of Physical Chemistry A}
  \textbf{2019}, \emph{123}, 1487--1497\relax
\mciteBstWouldAddEndPuncttrue
\mciteSetBstMidEndSepPunct{\mcitedefaultmidpunct}
{\mcitedefaultendpunct}{\mcitedefaultseppunct}\relax
\EndOfBibitem
\bibitem[Benali \latin{et~al.}(2020)Benali, Gasperich, Jordan, Applencourt,
  Luo, Bennett, Krogel, Shulenburger, Kent, Loos, \latin{et~al.}
  others]{benali2020toward}
Benali,~A.; Gasperich,~K.; Jordan,~K.~D.; Applencourt,~T.; Luo,~Y.;
  Bennett,~M.~C.; Krogel,~J.~T.; Shulenburger,~L.; Kent,~P.~R.; Loos,~P.-F.,
  \latin{et~al.}  Toward a systematic improvement of the fixed-node
  approximation in diffusion Monte Carlo for solids—A case study in diamond.
  \emph{The Journal of Chemical Physics} \textbf{2020}, \emph{153},
  184111\relax
\mciteBstWouldAddEndPuncttrue
\mciteSetBstMidEndSepPunct{\mcitedefaultmidpunct}
{\mcitedefaultendpunct}{\mcitedefaultseppunct}\relax
\EndOfBibitem
\bibitem[Shee \latin{et~al.}(2018)Shee, Arthur, Zhang, Reichman, and
  Friesner]{shee2018phaseless}
Shee,~J.; Arthur,~E.~J.; Zhang,~S.; Reichman,~D.~R.; Friesner,~R.~A. Phaseless
  auxiliary-field quantum Monte Carlo on graphical processing units.
  \emph{Journal of chemical theory and computation} \textbf{2018}, \emph{14},
  4109--4121\relax
\mciteBstWouldAddEndPuncttrue
\mciteSetBstMidEndSepPunct{\mcitedefaultmidpunct}
{\mcitedefaultendpunct}{\mcitedefaultseppunct}\relax
\EndOfBibitem
\bibitem[Sharma \latin{et~al.}(2017)Sharma, Holmes, Jeanmairet, Alavi, and
  Umrigar]{ShaHolUmr}
Sharma,~S.; Holmes,~A.~A.; Jeanmairet,~G.; Alavi,~A.; Umrigar,~C.~J.
  Semistochastic Heat-Bath Configuration Interaction Method: Selected
  Configuration Interaction with Semistochastic Perturbation Theory. \emph{J.
  Chem. Theory Comput.} \textbf{2017}, \emph{13}, 1595--1604, PMID:
  28263594\relax
\mciteBstWouldAddEndPuncttrue
\mciteSetBstMidEndSepPunct{\mcitedefaultmidpunct}
{\mcitedefaultendpunct}{\mcitedefaultseppunct}\relax
\EndOfBibitem
\bibitem[Smith \latin{et~al.}(2017)Smith, Mussard, Holmes, and
  Sharma]{smith2017cheap}
Smith,~J.~E.; Mussard,~B.; Holmes,~A.~A.; Sharma,~S. Cheap and near exact
  CASSCF with large active spaces. \emph{J. Chem. Theory Comput.}
  \textbf{2017}, \emph{13}, 5468--5478\relax
\mciteBstWouldAddEndPuncttrue
\mciteSetBstMidEndSepPunct{\mcitedefaultmidpunct}
{\mcitedefaultendpunct}{\mcitedefaultseppunct}\relax
\EndOfBibitem
\bibitem[Mahajan and Sharma(2020)Mahajan, and Sharma]{mahajan2020efficient}
Mahajan,~A.; Sharma,~S. Efficient local energy evaluation for multi-Slater wave
  functions in orbital space quantum Monte Carlo. \emph{The Journal of Chemical
  Physics} \textbf{2020}, \emph{153}, 194108\relax
\mciteBstWouldAddEndPuncttrue
\mciteSetBstMidEndSepPunct{\mcitedefaultmidpunct}
{\mcitedefaultendpunct}{\mcitedefaultseppunct}\relax
\EndOfBibitem
\bibitem[Rom \latin{et~al.}(1997)Rom, Charutz, and Neuhauser]{rom1997shifted}
Rom,~N.; Charutz,~D.; Neuhauser,~D. Shifted-contour auxiliary-field Monte
  Carlo: circumventing the sign difficulty for electronic-structure
  calculations. \emph{Chemical physics letters} \textbf{1997}, \emph{270},
  382--386\relax
\mciteBstWouldAddEndPuncttrue
\mciteSetBstMidEndSepPunct{\mcitedefaultmidpunct}
{\mcitedefaultendpunct}{\mcitedefaultseppunct}\relax
\EndOfBibitem
\bibitem[L{\"o}wdin(1955)]{lowdin1955quantum}
L{\"o}wdin,~P.-O. Quantum theory of many-particle systems. I. Physical
  interpretations by means of density matrices, natural spin-orbitals, and
  convergence problems in the method of configurational interaction.
  \emph{Physical Review} \textbf{1955}, \emph{97}, 1474\relax
\mciteBstWouldAddEndPuncttrue
\mciteSetBstMidEndSepPunct{\mcitedefaultmidpunct}
{\mcitedefaultendpunct}{\mcitedefaultseppunct}\relax
\EndOfBibitem
\bibitem[Balian and Brezin(1969)Balian, and Brezin]{balian1969nonunitary}
Balian,~R.; Brezin,~E. Nonunitary Bogoliubov transformations and extension of
  Wick’s theorem. \emph{Il Nuovo Cimento B (1965-1970)} \textbf{1969},
  \emph{64}, 37--55\relax
\mciteBstWouldAddEndPuncttrue
\mciteSetBstMidEndSepPunct{\mcitedefaultmidpunct}
{\mcitedefaultendpunct}{\mcitedefaultseppunct}\relax
\EndOfBibitem
\bibitem[Sun \latin{et~al.}(2018)Sun, Berkelbach, Blunt, Booth, Guo, Li, Liu,
  McClain, Sayfutyarova, Sharma, Wouters, and Chan]{sun2018pyscf}
Sun,~Q.; Berkelbach,~T.~C.; Blunt,~N.~S.; Booth,~G.~H.; Guo,~S.; Li,~Z.;
  Liu,~J.; McClain,~J.~D.; Sayfutyarova,~E.~R.; Sharma,~S.; Wouters,~S.;
  Chan,~K.-L.~G. PySCF: the Python-based simulations of chemistry framework.
  \emph{WIREs Comput. Mol. Sci.} \textbf{2018}, \emph{8}, e1340\relax
\mciteBstWouldAddEndPuncttrue
\mciteSetBstMidEndSepPunct{\mcitedefaultmidpunct}
{\mcitedefaultendpunct}{\mcitedefaultseppunct}\relax
\EndOfBibitem
\bibitem[dqm()]{dqmc_code}
\url{https://github.com/sanshar/VMC/}\relax
\mciteBstWouldAddEndPuncttrue
\mciteSetBstMidEndSepPunct{\mcitedefaultmidpunct}
{\mcitedefaultendpunct}{\mcitedefaultseppunct}\relax
\EndOfBibitem
\bibitem[afq()]{afqmc_files}
\url{https://github.com/ankit76/ph_afqmc}\relax
\mciteBstWouldAddEndPuncttrue
\mciteSetBstMidEndSepPunct{\mcitedefaultmidpunct}
{\mcitedefaultendpunct}{\mcitedefaultseppunct}\relax
\EndOfBibitem
\bibitem[Buonaura and Sorella(1998)Buonaura, and
  Sorella]{buonaura1998numerical}
Buonaura,~M.~C.; Sorella,~S. Numerical study of the two-dimensional Heisenberg
  model using a Green function Monte Carlo technique with a fixed number of
  walkers. \emph{Physical Review B} \textbf{1998}, \emph{57}, 11446\relax
\mciteBstWouldAddEndPuncttrue
\mciteSetBstMidEndSepPunct{\mcitedefaultmidpunct}
{\mcitedefaultendpunct}{\mcitedefaultseppunct}\relax
\EndOfBibitem
\bibitem[Motta \latin{et~al.}(2017)Motta, Ceperley, Chan, Gomez, Gull, Guo,
  Jim{\'e}nez-Hoyos, Lan, Li, Ma, \latin{et~al.} others]{motta2017towards}
Motta,~M.; Ceperley,~D.~M.; Chan,~G. K.-L.; Gomez,~J.~A.; Gull,~E.; Guo,~S.;
  Jim{\'e}nez-Hoyos,~C.~A.; Lan,~T.~N.; Li,~J.; Ma,~F., \latin{et~al.}  Towards
  the solution of the many-electron problem in real materials: Equation of
  state of the hydrogen chain with state-of-the-art many-body methods.
  \emph{Physical Review X} \textbf{2017}, \emph{7}, 031059\relax
\mciteBstWouldAddEndPuncttrue
\mciteSetBstMidEndSepPunct{\mcitedefaultmidpunct}
{\mcitedefaultendpunct}{\mcitedefaultseppunct}\relax
\EndOfBibitem
\bibitem[Trail and Needs(2017)Trail, and Needs]{trail2017shape}
Trail,~J.~R.; Needs,~R.~J. Shape and energy consistent pseudopotentials for
  correlated electron systems. \emph{The Journal of chemical physics}
  \textbf{2017}, \emph{146}, 204107\relax
\mciteBstWouldAddEndPuncttrue
\mciteSetBstMidEndSepPunct{\mcitedefaultmidpunct}
{\mcitedefaultendpunct}{\mcitedefaultseppunct}\relax
\EndOfBibitem
\bibitem[Whitlock \latin{et~al.}(1979)Whitlock, Ceperley, Chester, and
  Kalos]{whitlock1979properties}
Whitlock,~P.; Ceperley,~D.; Chester,~G.; Kalos,~M. Properties of liquid and
  solid He 4. \emph{Physical Review B} \textbf{1979}, \emph{19}, 5598\relax
\mciteBstWouldAddEndPuncttrue
\mciteSetBstMidEndSepPunct{\mcitedefaultmidpunct}
{\mcitedefaultendpunct}{\mcitedefaultseppunct}\relax
\EndOfBibitem
\bibitem[Ceperley and Kalos(1986)Ceperley, and Kalos]{ceperley1986quantum}
Ceperley,~D.~M.; Kalos,~M. \emph{Monte Carlo methods in statistical physics};
  Springer, 1986; pp 145--194\relax
\mciteBstWouldAddEndPuncttrue
\mciteSetBstMidEndSepPunct{\mcitedefaultmidpunct}
{\mcitedefaultendpunct}{\mcitedefaultseppunct}\relax
\EndOfBibitem
\bibitem[Rothstein(2013)]{rothstein2013survey}
Rothstein,~S.~M. A survey on pure sampling in quantum Monte Carlo methods.
  \emph{Canadian Journal of Chemistry} \textbf{2013}, \emph{91}, 505--510\relax
\mciteBstWouldAddEndPuncttrue
\mciteSetBstMidEndSepPunct{\mcitedefaultmidpunct}
{\mcitedefaultendpunct}{\mcitedefaultseppunct}\relax
\EndOfBibitem
\bibitem[Motta and Zhang(2017)Motta, and Zhang]{motta2017computation}
Motta,~M.; Zhang,~S. Computation of ground-state properties in molecular
  systems: Back-propagation with auxiliary-field quantum Monte Carlo.
  \emph{Journal of chemical theory and computation} \textbf{2017}, \emph{13},
  5367--5378\relax
\mciteBstWouldAddEndPuncttrue
\mciteSetBstMidEndSepPunct{\mcitedefaultmidpunct}
{\mcitedefaultendpunct}{\mcitedefaultseppunct}\relax
\EndOfBibitem
\bibitem[Muenter(1975)]{muenter1975electric}
Muenter,~J. Electric dipole moment of carbon monoxide. \emph{Journal of
  Molecular Spectroscopy} \textbf{1975}, \emph{55}, 490--491\relax
\mciteBstWouldAddEndPuncttrue
\mciteSetBstMidEndSepPunct{\mcitedefaultmidpunct}
{\mcitedefaultendpunct}{\mcitedefaultseppunct}\relax
\EndOfBibitem
\bibitem[Lide(2004)]{lide2004crc}
Lide,~D.~R. \emph{CRC handbook of chemistry and physics}; CRC press, 2004;
  Vol.~85\relax
\mciteBstWouldAddEndPuncttrue
\mciteSetBstMidEndSepPunct{\mcitedefaultmidpunct}
{\mcitedefaultendpunct}{\mcitedefaultseppunct}\relax
\EndOfBibitem
\bibitem[Huggins \latin{et~al.}(2021)Huggins, O'Gorman, Rubin, Reichman,
  Babbush, and Lee]{Huggins2021Jun}
Huggins,~W.~J.; O'Gorman,~B.~A.; Rubin,~N.~C.; Reichman,~D.~R.; Babbush,~R.;
  Lee,~J. {Unbiasing Fermionic Quantum Monte Carlo with a Quantum Computer}.
  \emph{arXiv} \textbf{2021}, \relax
\mciteBstWouldAddEndPunctfalse
\mciteSetBstMidEndSepPunct{\mcitedefaultmidpunct}
{}{\mcitedefaultseppunct}\relax
\EndOfBibitem
\end{mcitethebibliography}

\providecommand{\latin}[1]{#1}
\makeatletter
\providecommand{\doi}
  {\begingroup\let\do\@makeother\dospecials
  \catcode`\{=1 \catcode`\}=2 \doi@aux}
\providecommand{\doi@aux}[1]{\endgroup\texttt{#1}}
\makeatother
\providecommand*\mcitethebibliography{\thebibliography}
\csname @ifundefined\endcsname{endmcitethebibliography}
  {\let\endmcitethebibliography\endthebibliography}{}

\end{document}